\documentclass[twocolumn]{aastex62}

\shorttitle{Radio observations of short-GRB host galaxies}
\shortauthors{Klose et al.}

\usepackage{xcolor}
\usepackage{enumerate}

\def\HII{[H\,{\sc II}]}
\def\OII{[O\,{\sc II}]}
\def\OIII{[O\,{\sc III}]}
\def\SII{[S\,{\sc II}]}

\def\NII{[N\,{\sc II}]}

\usepackage{hyperref}
\hypersetup{
breaklinks=true,
bookmarks=true,     
unicode=true,       
colorlinks=true,    
linkcolor=red,   
citecolor=blue,     
filecolor=red,      
urlcolor=black      
}

\newcommand{\swift}{\textit{Swift}}
\newcommand{\HST}{\textit{HST}}
\newcommand{\WISE}{\textit{WISE}\ }

\newcommand{\msun}{M$_\odot$}
\newcommand{\msunyr}{M$_\odot$\,yr$^{-1}$}

\def\kr{\hbox{ \raisebox{-1.0mm}{$\stackrel{<}{\sim}$} }}

\begin{document}


\title{Deep ATCA and VLA radio observations of short-GRB host galaxies.\\
Constraints on star-formation rates, afterglow flux, and
kilonova radio flares.}

\author{S.~Klose}
\author{A.~M.~Nicuesa Guelbenzu}
\affiliation{Th\"uringer Landessternwarte Tautenburg, Sternwarte 5, 07778 
  Tautenburg, Germany}

\author{M.~Micha{\l}owski}
\affiliation{Astronomical Observatory Institute, Faculty of Physics, Adam
  Mickiewicz University, ul. S\l oneczna 36, 60-286, Pozna\'n, Poland}
\affiliation{Scottish Universities Physics Alliance (SUPA), Institute for
  Astronomy, University of 
  Edinburgh, Royal Observatory, Blackford Hill, EH9 3HJ, Edinburgh, UK}

\author[0000-0001-9162-2371]{L.~K.~Hunt}
\affiliation{INAF -- Osservatorio Astrofisico di Arcetri, I-50125 Firenze, Italy}

\author{D.~H.~Hartmann}
\affiliation{Department of Physics and Astronomy, Clemson University, Clemson,
  SC 29634, USA}

\author{J.~Greiner}
\affiliation{Max-Planck-Institut f\"u r extraterrestrische Physik, D-85748 
  Garching, Germany}

\author{A.~Rossi}
\affiliation{INAF -- Osservatorio di Astrofisica e Scienza dello Spazio,
  via Piero Gobetti 93/3, 40129 Bologna, Italy}
\affiliation{INAF -- Osservatorio Astronomico di Roma, via Frascati 33, 00040
  Monte Porzio Catone, Italy}

\author{E.~Palazzi}
\affiliation{INAF -- Osservatorio di Astrofisica e Scienza dello Spazio,
  via Piero Gobetti 93/3, 40129 Bologna, Italy}

\author{S.~Bernuzzi}
\affiliation{Theoretisch-Physikalisches Institut, 
  Friedrich-Schiller-Universit{\"a}t Jena, 07743, Jena, Germany}  


\begin{abstract} 

We report the results of an extensive radio-continuum observing
campaign of host galaxies of short gamma-ray bursts (GRBs). The goal
of this survey was to search for optically obscured star formation,
possibly indicative of a population of young short-GRB progenitors.
Our sample comprises  the hosts and host-galaxy candidates of 16
short-GRBs from 2005 to 2015, corresponding to roughly 1/3 of the
presently known ensemble of well-localized short bursts.  Eight GRB
fields were observed with ATCA (at 5.5 and 9.0~GHz), and eight fields
with the VLA (mostly at 5.5~GHz). The observations typically achieved
a 1$\sigma_{\rm rms}$ of 5 to 8~$\mu$Jy. In most cases they were
performed years after the corresponding burst. No new short-GRB host
with optically obscured star formation was found. Only one host
galaxy was detected, the one of GRB 100206A at $z$=0.407.  However,
its starburst nature was already known from optical/IR data.  Its
measured radio flux can be interpreted as being due to a star
formation rate (SFR) of about 60\,\msunyr. This is in good agreement
with earlier expectations based on the observed broad-band spectral
energy distribution of this galaxy.  The 15 non-detections constrain
the SFRs of the suspected host galaxies and provide upper limits on
late-time luminosities of the associated radio afterglows and
predicted kilonova radio flares. The non-detection of radio
emission from GRB explosion sites confirms the intrinsically low
luminosity of short-GRB afterglows and places significant constraints
on the parameter space of magnetar-powered radio flares. Luminous
radio flares from fiducial massive magnetars have not been found.
 
\end{abstract}


\section{Introduction}

\label{Sect.Intro}
  
The last 20 years have seen a revolution in gamma-ray burst (GRB)
science, driven by new dedicated satellite missions, a deeper
theoretical understanding of stellar explosions, and substantial
efforts in sophisticated world-wide observing campaigns.  On the one
hand, long GRBs, which represent about 90\% of all well-localized
bursts, have been found to originate in the collapse of very massive
stars \citep[e.g.,][]{Hjorth2012grbu,WB2006,Cano2016}. On the other
hand, the link of short GRBs  to merging compact stars was
unambiguously confirmed by the gravitational wave event GW 170817
\citep[e.g.,][]{Abbott2017ApJ...848L..12A,Monitor:2017mdv,Coulter2017Sci...358.1556C,
  Kasen2017Natur.551...80K}.

In order to understand diversity in burst populations, host galaxy
studies are an important observational tool. Such studies have shown
that  long bursts appear to originate exclusively from late-type galaxies
\citep[e.g.,][]{Fruchter2006,Wainwright2007,Lyman2017},
with star-formation rates (SFRs) spanning a wide range from
$\sim$0.01 to $>$100\,\msunyr
\citep[e.g.,][]{Sokolov2001,Christensen2004,Castro2006,Castro2010,Fan2010,Hunt2011}.
Short bursts on the other hand arise from all morphological types of
galaxies
\citep[e.g.,][]{Fong2010ApJ...708....9F,Fong2013ApJ...769...56F,Berger2014ARAA}.
By the end of 2018, well-defined positions had been found for $\sim$50
short GRBs, allowing for an identification and investigation of their
hosts or host-galaxy {\it candidates} \citep[for a review,
  see][]{Berger2014ARAA}.  These observations show that  about 3/4 of
all short GRBs originated in star-forming galaxies
\citep[e.g.,][]{Leibler2010,Fong2010,Fong2013,Fong2013ApJ...769...56F},
with SFRs similar to what has been found for the hosts of long GRBs.

Star-formation rates in GRB host galaxies (long and short) are often
derived from measured emission line fluxes in optical bands
\citep[e.g.,][]{DAvanzo2009}. However, these lines can be affected by
extinction from cosmic dust. Radio observations on the other hand
trace synchrotron radiation from relativistic electrons originating
from supernova remnants. Since the supernova rate is 
directly related to the star formation rate,  radio-continuum
observations provide an unobscured view of the star-forming activity
in a galaxy over the last 10 to 100\,Myr
\citep[e.g.,][]{Greis2017MNRAS.470.489}.

Consequently, in recent years radio emission of  host galaxies of {\it long
GRBs} has been studied with the goal of deriving the unobscured SFR based on
the measured radio-continuum flux \citep[e.g.,][]{Berger2003ApJ588.Radio,
Stanway2010,Hatsukade2012,Michalowski2012,PerleyPerley.Radio.2013,
Perley2015}. In a
comprehensive analysis, \citet{Greiner2016A&A593A} list 61 long-GRB host
galaxies at redshifts $0.01 \kr z \kr 2.5$  that have been observed in the
radio band. Among these, 18 were detected and 12 (i.e., $\sim$20\%) have
a radio-derived SFR larger than 20\,\msunyr.

Given these results for long-GRB host galaxies,  the identification of
three star-bursting hosts among the relatively small short-GRB
ensemble \citep[GRB 100206A,  GRB 120804A,  GRB
071227:][respectively]{Perley2012,Berger2013.765,Nic2014ApJ}
deserves attention. More such cases could be indicative of a
population of young merger systems (neutron star - neutron star,
neutron star - black hole)  associated with short GRBs, as
predicted by some stellar population synthesis models
(\citealt{Voss2003MNRAS,Belczynski2006,Belczynski2007,Mapelli2018MNRAS.481.5324M,
  Kruckow2018MNRAS.481.1908K,Belczynski2018}). Indeed,
a general existence of young merger
systems has been  observationally supported by the discovery of the
{\it galactic} NS-NS binary PSR~J0737$-$3039  which will merge
within only $\sim$85~Myr
(\citealt{Burgay2003Natur,Tauris2017ApJ...846..170T}). 

With this motivation in mind, we describe the results of a comprehensive
radio-continuum study  of star-forming hosts (and host-galaxy
candidates; e.g., \citealt{Giacomazzo2013,Berger2014ARAA}) of 16
short GRBs.\footnote{Host-galaxy candidates have no secure
association with the burst under consideration but are selected
according to a chance coincidence analysis given their apparent
magnitude in a certain photometric band and their angular offset
from the  burst position \citep{Bloom2002}.}  This  is roughly 1/3
of all well-localized short bursts. The observations were performed
using NSF's {\it Karl G. Jansky} Very Large Array (VLA) and the
Australia Telescope Compact Array (ATCA). 

In the following we adopt a flat $\Lambda$CDM cosmology with Hubble
constant H$_0=68$\,km\,s$^{-1}$\,Mpc$^{-1}$ and density parameters
$\Omega_M=0.31$ and $\Omega_\Lambda=0.69$ \citep{Planck2016}.  For the
observed flux density $F_\nu$ we use the convention  $F_\nu \sim
\nu^{-\beta}$, where $\nu$ is the frequency and $\beta$ is the
spectral slope (in energy space, not photon number space). 

\section{Target selection}

\label{Sect:Targets}
  
Short GRBs were selected according to the compilations in
\cite{Kann2011} and  \cite{Berger2014ARAA}, supplemented by the 
database maintained by
one of us (J.G.)\footnote{http://www.mpe.mpg.de/$\sim$jcg/grbgen.html.} One
controversial case was added to our list of short bursts (GRB
100816A, $T_{90}$ = 2.9$\pm$0.6~s; see Appendix).

Following the approach in \cite{Nic2014ApJ,Nic2015}, in our radio
survey we focused on a search for optically obscured star-forming
activity in short-GRB hosts. We excluded events securely
categorized as having early-type hosts since in these cases we did
not expect a detectable radio emission due to star-forming activity.
In doing so, our first reference concerning the host-galaxy
classification was the list of short-GRB host-galaxies provided by
\cite{Berger2014ARAA}, which containes events until mid 2013 (his
table 2). We updated this list if additional imaging data were
available that required to re-consider the original classification
of a host (see Appendix). Concerning short GRBs not listed in
\cite{Berger2014ARAA}, we  made use of publicely available  data or
performed our own observations in order to classify suspected host
galaxies.  Cases with no clear host-galaxy classification (early or
late-type galaxy) were not rejected. This concerned two events (GRB
090621B and 101224A). Once this step was done our first choice
were short-GRB hosts  that were detected in the infrared bands by the
Wide-field Infrared Survey Explorer ({\it WISE}) satellite
\citep{Wright2010} but which are not  elliptical galaxies.  Cases with
more than one host-galaxy candidate were included in our target list
as long as at least one candidate is not an elliptical galaxy. Radio
observations could in principle reveal if one host-galaxy candidate is
special in some sense, e.g., exhibits  a high SFR.

\WISE observed the entire sky in four bands at 3.4, 4.6, 12, and 22$\mu$m
\citep[W1, W2, W3, W4, see][]{Wright2010}. The \WISE
catalog\footnote{http://wise.ssl.berkeley.edu/} lists all sources with a
measured signal-to-noise ratio greater than 5 in at least one band.  Even
though the catalog does not have the same sensitivity in all directions, for
our purpose it was the best available infrared all-sky data base. 

Our analysis showed that 12 short-GRB host galaxies were
detected by \WISE in at  least one band
(Table~\ref{tab:short_WISE}). This is $\sim$1/3 of the sample of
short-GRB hosts (by the year 2015). No host galaxy has been detected
at the longest wavelength (W4), while five hosts  were detected in the
W1, W2 and W3 bands: GRB 060502B at $z$=0.287, GRB 071227 at
$z$=0.381, GRB 100206A at $z$=0.407, GRB 150101B at $z$=0.134, and GRB
161104A with an unknown $z$. It is noteworthy that these are not the
cosmologically nearest short-GRB hosts. According to their position in
the \WISE color-color diagram,  which combines the W1-W2 color with
the W2-W3 color \citep[][their figure 12]{Wright2010}, the hosts of
GRB 071227 and 100206A fall into the starburst category. Similarly,
the host of GRB 060502B might be a spiral, while the hosts of GRB
150101B  and 161104A could be elliptical galaxies. \citep[see
  also][]{2016ApJ...833..151F, Fong2016GCN.20168....1F, Xie2016}.
Moreover, the host of GRB 150101B is the first GRB host galaxy that
harbors an AGN \citep{Xie2016}.

A subset of six host galaxies (or host-galaxy candidates) were detected
by \WISE in W1 and W2 but not in W3 and W4, while two hosts have only a
detection in filter W1. The former group includes GRB 050724, 070724, 070729,
111222A, 130603B, and  150424A, the latter group GRB 070429B and
100816A. Among these, the field of GRB 070729 contains several host-galaxy
candidates. Unfortunately, no redshift is known in this case.

\begin{deluxetable*}{lllllllll}
\tablecolumns{9} 
\tablewidth{0pt} 
\tablecaption{Host galaxies of short GRBs (by 2015) in the \WISE satellite archive (all-sky).}
\tablehead{
\colhead{GRB} & 
\colhead{Decl.} & 
\colhead{W1} & 
\colhead{W2} & 
\colhead{W3} & 
\colhead{W4} & 
\colhead{W1-W2} & 
\colhead{W2-W3} & 
\colhead{gal. type} \\
\colhead{(1)} &
\colhead{(2)} &
\colhead{(3)} &
\colhead{(4)} &
\colhead{(5)} &
\colhead{(6)} &
\colhead{(7)} &
\colhead{(8)} &
\colhead{(9)} }
\startdata
 050724   & $-$27 & 15.35$\,\pm\,$0.05 & 15.41$\,\pm\,$0.13 & $>$12.37           & $>$8.42 &$-$0.06$\,\pm\,$ 0.14  &$<$3.04 $\,\pm\,$ 0.13 \\             
 060502B  & +52   & 15.29$\,\pm\,$0.03 & 14.94$\,\pm\,$0.04 & 12.88$\,\pm\,$0.35 & $>$9.65 &   0.35$\,\pm\,$ 0.05  &  2.06 $\,\pm\,$ 0.35 & Spiral\\     
 070429B  & $-$32 & 17.38$\,\pm\,$0.15 &     $>$16.68       & $>$12.44           & $>$9.01 &$<$0.70$\,\pm\,$ 0.15  &  --                  \\[1mm]        
 070724   & $-$18 & 17.03$\,\pm\,$0.11 & 16.37$\,\pm\,$0.22 & 12.34$\,\pm\,$0.37 & $>$9.13 &   0.66$\,\pm\,$ 0.25  &  4.03 $\,\pm\,$ 0.43 \\             
 070729   & $-$39 & 16.54$\,\pm\,$0.05 & 16.69$\,\pm\,$0.18 & $>$13.22           & $>$9.34 &$-$0.15$\,\pm\,$ 0.19  &$<$3.47 $\,\pm\,$ 0.18 \\             
 071227   & $-$55 & 15.58$\,\pm\,$0.03 & 15.15$\,\pm\,$0.05 & 11.88$\,\pm\,$0.16 & $>$9.70 &   0.43$\,\pm\,$ 0.06  &  3.27 $\,\pm\,$ 0.17 & LIRG\\[1mm]  
 100206A  & +13   & 15.71$\,\pm\,$0.05 & 15.15$\,\pm\,$0.10 & 11.31$\,\pm\,$0.18 & $>$8.57 &   0.56$\,\pm\,$ 0.11  &  3.84 $\,\pm\,$ 0.21 & LIRG\\       
 100816A  & +26   & 17.08$\,\pm\,$0.12 & 16.87$\,\pm\,$0.36 & $>$12.05           & $>$8.79 &   0.21$\,\pm\,$ 0.38  &$<$4.82 $\,\pm\,$ 0.36 \\             
 111222A  & +65   & 13.26$\,\pm\,$0.02 & 13.16$\,\pm\,$0.03 & $>$12.45           & $>$8.91 &   0.10$\,\pm\,$ 0.04  &$<$0.71 $\,\pm\,$ 0.03 \\[1mm]        
 130603B  & +17   & 17.05$\,\pm\,$0.11 & 16.95$\,\pm\,$0.37 & $>$12.53           & $>$8.91 &   0.10$\,\pm\,$ 0.39  &$<$4.42 $\,\pm\,$ 0.37 \\             
 150101B  & $-$10 & 13.43$\,\pm\,$0.03 & 13.18$\,\pm\,$0.03 & 12.39$\,\pm\,$0.48 & $>$8.78 &   0.25$\,\pm\,$ 0.04  &  0.79 $\,\pm\,$ 0.48 & Elliptical\\ 
 150424A  & $-$26 & 16.44$\,\pm\,$0.07 & 16.10$\,\pm\,$0.17 & $>$12.12           & $>$8.69 &   0.34$\,\pm\,$ 0.18  &$<$3.98 $\,\pm\,$ 0.17 \\[1mm]        
\enddata 
\tablecomments{Listed here are all short-GRB hosts and host-galaxy
  candidates  with detections in at least one of the four \WISE bands
  (\citealt{Wright2010}).  Upper limits are given according to the \WISE
  catalog.  Column \#2 gives the Declination (J2000) in degrees (which decides
  between a potential ATCA or VLA target), columns  \#3 to \#6 the magnitudes
  on the four \WISE bands, and columns  \#7 and \#8 the corresponding W1-W2
  and W2-W3 color (in mag), respectively, which enters the \WISE color-color
  diagram and provides the type of the galaxy (column \#9; LIRG =  Luminous
  Infrared Galaxy).  All magnitudes refer to the Vega photometric system.} 
\label{tab:short_WISE}
\end{deluxetable*}

Additional targets were selected according to redshift, because for
sensitivity considerations, we wanted to include nearby hosts if possible;
thus an additional group of 5 targets entered our list  because of their
redshift ($z<$0.5), even though they were not seen by \WISE in any band (GRB
061006, 061201, 061210A, 070809, and 080123). 

Finally, we observed three additional  short-GRB hosts with unknown redshifts (GRB
090621B, 101224A, and 130515A). The common trait of these hosts is that there
was no optically-detected transient.  The status of the identification of
their host galaxies is thus very different, ranging from ''no known
host-galaxy candidate'' (GRB 090621B) to ''a host-galaxy candidate outside the
\swift/XRT error circle'' (GRB 130515A). If they were dusty star-bursting
galaxies similar to  the host of the short GRB 120804A
\citep[SFR$\sim$300\,\msunyr;][]{Berger2013.765}, we would have been able to
detect them up to a redshift $z \sim 1.5$. 

Altogether, our sample consists of 16 short-GRB host  galaxies which span a
redshift range from $z$=0.1 to 0.8.
According to \cite{Bromberg2013} (see also \citealt{Wanderman2015}),
bursts with $T_{90}<$0.8~s represent a much cleaner short-burst sample
than longer short-GRB events. Most of our targets belong to this
$T_{90}<$0.8~s category, i.e.,
based on their duration these events are confidently a merger and not a
collapsar.

\begin{deluxetable*}{lclr rcr}
\tablecolumns{7} 
\tablewidth{0pt} 
\tablecaption{Target list.}
\tablehead{
  \colhead{GRB}  &
  \colhead{select.}&
  \colhead{$z$}  &
  \colhead{R.A. (J2000)}&
  \colhead{Decl.}&
  \colhead{position} &
  \colhead{$T_{90}$[s]}\\
\colhead{(1)} &
\colhead{(2)} &
\colhead{(3)} &
\colhead{(4)} &
\colhead{(5)} &
\colhead{(6)} &
\colhead{(7)} }
\startdata
 {\bf ATCA}        &        &              &                &           &                \\[1mm] 
 050724  &WISE      & 0.258  &  16:24:44.36 &  $-$27:32:27.5 &  OT       &  $3.0\pm1.0$  \\
 061006  &redshift  & 0.4377 &  07:24:07.66 &  $-$79:11:55.1 &  OT       &  0.42         \\
 061201  &redshift  & 0.111  &  22:08:32.09 &  $-$74:34:47.1 &  OT       &  $0.8\pm0.1$  \\
 070729  &WISE      &        &  03:45:15.97 &  $-$39:19:20.5 &  XRT      &  $0.9\pm0.1$  \\[1mm]
 070809  &redshift  & 0.2187 &  13:35:04.55 &  $-$22:08:30.8 &  OT       &  $1.3\pm0.1$   \\
 080123  &redshift  & 0.496  &  22:35:46.33 &  $-$64:54:02.7 &  XRT      &     $0.4$      \\
 130515A &noOT      &        &  18:53:45.71 &  $-$54:16:45.5 &  XRT      & $0.29\pm0.06$    \\
 150424A &WISE      & 0.2981 &  10:09:13.38 &  $-$26:37:51.5 &  OT       & 0.4            \\[3mm]
 {\bf VLA}             &        &              &                &                           \\[1mm] 
 060502B &WISE      & 0.287  &  18:35:44.97 &     52:37:52.4 &  XRT      & $0.09\pm0.02$  \\
 061210  &redshift  & 0.4095 &  09:38:05.17 &     15:37:17.5 &  XRT      & $0.047$        \\ 
 070724A &WISE      & 0.457  &  01:51:14.07 &  $-$18:35:39.3 &  OT       & $0.4\pm0.04$   \\
 090621B &noOT      &        &  20:53:53.16 &     69:01:41.5 &  XRT      & $0.14\pm0.04$  \\[1mm] 
 100206A &WISE      & 0.4068 &  03:08:39.00 &     13:09:24.8 &  XRT      & $0.12\pm0.03$  \\ 
 100816A &WISE      & 0.8049 &  23:26:57.56 &     26:34:42.9 &  OT       & $2.9\pm 0.6$   \\
 101224A &noOT      &        &  19:03:41.72 &     45:42:49.5 &  XRT      & $0.20\pm0.01$  \\ 
 130603B &WISE      & 0.3565 &  11:28:48.15 &     17:04:18.0 &  OT       & $0.18\pm0.02$  \\ 
\enddata 
\tablecomments{Column \#2 classifies the used selection criteria: (i) WISE all-sky
  survey, (ii) small redshift, or (iii) no detected optical transient
  (noOT; see Sect.~\ref{Sect:Targets}).  Column \#3 provides the
  redshift (see Appendix), if available. Columns \#4 and \#5  either
  give the position of the OT or  the central coordinates of the  XRT
  error circle  ({http://www.swift.ac.uk/xrt$_-$positions/};
  \citealt{Goad2007,Evans2009}) as of January 2019.  Column \#6
  distinguishes between the OT and the XRT position of the afterglow.
  Column \#7 provides the burst duration $T_{90}$, mostly measured in
  the \swift/BAT energy band (15-350 keV). For references see the
  Appendix.}
\label{Tab:targetlist}
\end{deluxetable*}

\section{Observations and data reduction}

\subsection{Radio data}

All radio-continuum observations were performed with ATCA and VLA in the
years 2013 to 2015.

\begin{deluxetable*}{lcl rcccc r}
\tablecolumns{9} 
\tablewidth{0pt} 
\tablecaption{Summary of the radio observations.}
\tablehead{
\colhead{GRB} & 
\colhead{date obs.} & 
\colhead{config.} & 
\colhead{time} & 
\colhead{flux calib.} & 
\colhead{phase calib.} & 
\colhead{freq.} & 
\colhead{beam size} & 
\colhead{1$\sigma_{\rm rms}$} \\
\colhead{(1)} &
\colhead{(2)} &
\colhead{(3)} &
\colhead{(4)} &
\colhead{(5)} &
\colhead{(6)} &
\colhead{(7)} &
\colhead{(8)} &
\colhead{(9)}}
\startdata
{\bf ATCA}   &                 &     &         &            &             &        &                      &     \\[1mm]
     050724  &  23-10-2015     & 6A  &  10.4   & 0823--500  & 1622--253   & 5.5    & 4.4 $\,\times\,$ 1.5 & 5.0 \\
             &&&&&&                                                         9.0    & 2.7 $\,\times\,$ 0.9 & 5.2 \\
     061006  &  20-07-2013     & 6A  &  7.0    & 1934--638  & 0637--752   & 5.5    & 3.0 $\,\times\,$ 1.3 & 7.8 \\
             &&&&&&                                                         9.0    & 1.9 $\,\times\,$ 0.8 & 9.3 \\
     061201  &  21-07-2013     & 6A  &  11.5   & 0823--500  & 2142--758   & 5.5    & 2.3 $\,\times\,$ 1.5 & 5.2 \\  
             &&&&&&                                                         9.0    & 1.4 $\,\times\,$ 0.9 & 5.2 \\    
             &  26-07-2013     & 6A  &  3.2    & 1934--638  & 2142--758   & 5.5,9.0&                      &     \\     
             &  25-07-2013     & 6A  &  0.7    & 1934--638  & 2142--758   & 5.5,9.0&                      &     \\
     070729  &  28-04-2015     & 6A  &  11.0   & 0823--500  & 1934--638   & 5.5    & 4.5 $\,\times\,$ 2.5 & 7.3 \\
             &&&&&&                                                         9.0    & 2.7 $\,\times\,$ 1.5 & 6.9 \\
     070809  &  19-06-2013     & 6A  &  9.5    & 1934--638  & 1308--220   & 5.5    & 7.7 $\,\times\,$ 1.7 & 6.0 \\ 
             &&&&&&                                                         9.0    & 4.7 $\,\times\,$ 1.1 & 6.4 \\ 
     080123  &  25-07-2013     & 6A  &  8.7    & 1934--638  & 2353--686   & 5.5    & 2.6 $\,\times\,$ 1.5 & 6.4 \\
             &&&&&&                                                         9.0    & 1.6 $\,\times\,$ 0.9 & 6.6 \\
     130515A &  26-10-2015     & 6A  &  7.1    & 1934--638  & 1824--582   & 5.5    & 3.6 $\,\times\,$ 1.3 & 6.2 \\
             &&&&&&                                                         9.0    & 2.2 $\,\times\,$ 0.8 & 7.3 \\
     150424A &  19-06-2015     & 6D  &  1.8    & 1934--638  & 1034--293   & 5.5    &12.0 $\,\times\,$ 1.5 & 7.1 \\
             &&&&&&                                                         9.0    & 7.0 $\,\times\,$ 0.9 & 8.0 \\
             &  23-10-2015     & 6A  &  3.2    & 0823--500  & 1034--293   & 5.5,9.0&                      &     \\[3mm]
{\bf VLA}    &                 &     &         &            &             &        &                      &     \\[1mm]
     060502B &  18-01-2014     &  B  & 0.75    & 3C286      &  J1829+4844 & 3.0    & 3.0  $\,\times\,$ 2.0 & 8.8 \\
             &  09-10-2014     &  C  & 1.0     & 3C286      &  J1740+5211 & 5.5    & 4.0 $\,\times\,$ 3.4 & 6.4 \\
     061210  &  16-10-2014     &  C  & 1.0     & 3C286      &  J0854+2006 & 5.5    & 3.7 $\,\times\,$ 3.2 & 5.3 \\
     070724A &  19-09-2014     & DnC & 1.5     & 3C147      &  J0204--1701& 5.5    &11.2 $\,\times\,$ 7.0 & 7.6 \\
     090621B &  08-10-2015     &  D  & 0.75    & 3C286      &  J2022+6136 & 5.5    &16.2 $\,\times\,$10.6 & 5.7 \\
             &  17-10-2015     &  D  & 0.5     & 3C286      &  J2022+6136 & 5.5    &                        &     \\ 
             &  24-10-2015     &  D  & 0.5     & 3C286      &  J2022+6136 & 5.5    &                        &     \\ 
             &  28-10-2015     &  D  & 0.5     & 3C147      &  J2022+6136 & 5.5    &                        &     \\ 
     100206A &  10-10-2014     &  C  & 1.5     & 3C147      &  J0318+1628 & 5.5    & 4.0 $\,\times\,$ 3.9 & 13\\ 
     100816A &  09-10-2014     &  C  & 1.5     & 3C147      &  J2340+2641 & 5.5    & 5.8 $\,\times\,$ 3.8 & 5.8 \\
     101224A &  12-10-2015     &  D  & 0.5     & 3C286      &  J1845+4007 & 5.5    & 11.2 $\,\times\,$10.0 & 6.4 \\ 
             &  14-10-2015     &  D  & 0.75    & 3C48       &  J1845+4007 & 5.5    &                        &     \\ 
             &  20-10-2015     &  D  & 0.75    & 3C286      &  J1845+4007 & 5.5    &                        &     \\ 
             &  27-10-2015     &  D  & 0.75    & 3C48       &  J1845+4007 & 5.5    &                        &     \\ 
             &  30-10-2015     &  D  & 0.75    & 3C286      &  J1845+4007 & 5.5    &                        &     \\ 
     130603B &  13-11-2014     &  C  & 1.5     & 3C286      &  J1120+1420 & 5.5    & 3.9 $\,\times\,$ 3.5 & 6.5 \\
\enddata 
\tablecomments{Column \#2 provides the date of the start of the observations,
  column \#3 gives the telescope configuration.
  Column \#4 provides the telescope time
  (in hours). Columns \#5 and \#6 list the flux and phase calibrators,
  column \#7 the effective frequency. The
  last two columns contain the size of the synthesized beam in units of
  arcsec and the 1$\sigma$ rms of the resulting image (in units of $\mu$Jy
  beam$^{-1}$; measured using the \texttt{imstat} task under MIDAS and
  CASA, respectively). 
  Beam sizes and the rms refer to a robust parameter of 0.5. 
  Except for the S-band (3~GHz) observations of the host of GRB 060502B on
  18-01-2014, all VLA obervations were performed in the C band (effectively
  at 5.5~GHz). For GRB 061201, 090621B, 101224A, and 150424A the 1$\sigma$ rms
  and beam size refer to the combined data set.}
\label{Tab:obs.summary}
\end{deluxetable*}

ATCA observations were carried out in the 5.5 and 9.0~GHz bands (corresponding
to 6 and 3~cm, respectively), using the upgraded Compact Array Broadband
Backend (CABB) detector \citep{Wilson2011}  and all six 22-m antennae with the
6~km baseline (Table~\ref{Tab:obs.summary}; programme ID C2840, PI: A. Nicuesa
Guelbenzu). We set a technical  observing constraint of Decl. (J2000) $<
-15^\circ$.  ATCA can achieve a sensitivity at 5.5\,GHz of $\sigma_{\rm rms}
\sim 5\mu$Jy beam$^{-1}$ in a $\sim$10-hour integration.  Therefore, whenever
possible we stayed on a target between 10 and 12 hours in order to obtain the
deepest sensitivity, as well as the best coverage of the $uv$ plane for image
fidelity. The 5.5~GHz band is a good compromise between sensitivity and
expected radio flux on the one hand, and relatively lower Radio Frequency
Interference (RFI) on the other.  In the observing mode used here, CABB
integrates at 5.5 and 9.0\,GHz simultaneously, each with 2048 spectral
channels of width 1.0\,MHz.

Observations with the VLA went similarly deep as those with ATCA, with
integration times  typically a factor of 10 smaller because of the larger
number of antennas. Observations were performed via dynamical scheduling
(programme ID 13B-313, 14B-201, 15B-214; PI: A. Nicuesa Guelbenzu). Most data
were obtained in the C band (effectively at 5.5 GHz)  in C or D telescope
array configuration. One target was also observed in the S band in B
configuration (effectively at 3.0 GHz; Table~\ref{Tab:obs.summary}).  
All VLA observations used the WIDAR correlator with the wideband
setup and the 8-bit samplers. They were performed in 16 spectral
windows with a bandwidth of 128 MHz each, providing a total bandwidth of
2048 MHz per polarization.

In principle, observations in the 2.1~GHz band (16~cm) could provide a better
signal-to-noise ratio, since for star-forming galaxies the radio flux
typically scales as $F_\nu \propto \nu^{-\beta}$, $\beta\sim$0.7
\citep[e.g.,][]{Gioia1982,Tabatabaei2017,Klein2018AA...611A..55K}.  However,
this band is more affected by RFI. In addition, source crowding at 2.1~GHz is
higher, making the identification of faint radio sources more challenging.
Moreover, compared to observations in the 2.1~GHz band, at 5.5~GHz the radio
sky is less populated with bright radio sources which can also affect the
image reconstruction. 

During the observations,  bandpass and flux calibration were performed in the
usual manner. For ATCA observations  in most cases  the bright radio source
PKS\,1934--638 (R.A., Decl. J2000 = 19:39:25.026, $-$63:42:45.63) was used  as
the calibrator. If this source was not observable, then 0823--500 was observed
(R.A., Decl. J2000 = 08:25:26.869, $-$50:10:38.49). For VLA, the bright radio
sources 3C286, 3C147, and 3C48 were observed.  Nearby calibrators for phase
referencing  were chosen depending on the target coordinates via the
corresponding web interface provided by the ATCA and the VLA operating
institutes CSIRO and NRAO, respectively\footnote{For the coordinates and
fluxes of the ATCA phase calibrators see
https://www.narrabri.atnf. csiro.au/calibrators/c007/atcat.html.}.

Data reduction was performed with the standard software packages.
ATCA data were reduced using the Multichannel Image Reconstruction, Image Analysis and
Display (MIRIAD) software  package version 1.5
\citep{Sault1995}; for VLA we used the data reduction pipeline under CASA
version 5.1.2-4.
During the data processing, RFI was examined and eliminated as well as
possible from the data. All data were Fourier-transformed using the Briggs
robust weighting option \citep{Briggs1995}, varying this parameter between
0.5, 1.0, and 2.0, and selecting the one that gave the best compromise between
the resolution and the noise.

Usually targets were observed for $\sim 7 - 11$\,hr (ATCA) and $1-2$\,hr
(VLA) of elapsed telescope time.  Consequently, all observations are fairly
deep, typically reaching a 1$\sigma_{\rm rms}$ between 5 and 8~$\mu$Jy
beam$^{-1}$ (measured in an empty field around the target position). For 11 of
the 16 targets the resulting synthesized beam size was smaller than
5$\times\,$5 arcsec$^2$. Observing parameters are reported in
Table~\ref{Tab:obs.summary}.

\subsection{Optical data} 

Imaging:\ Optical/near-infrared (NIR) data was downloaded from public
data archives (ESO/VLT, Gemini, GTC, HST). Multi-color data were obtained
using the optical/near-infrared seven-channel imager GROND mounted at the 2.2m
ESO/MPG telescope on ESO/La Silla (Chile)
\citep{Greiner2007Msngr,Greiner2008}. In two cases (GRB 100816A, 150424A), the
optical SFR was also estimated based on a \texttt{Le
Phare}\footnote{http://www.cfht.hawaii.edu/~arnouts/lephare.html}
\citep[][]{Arnouts1999,Ilbert2006,Arnouts2013} analysis of the broad-band
spectral energy distribution (SED) of the galaxy.

Spectroscopy:\  In the majority of cases the optical SFR was
taken from the literature and is based on emission-line
spectroscopy. In two cases the optical SFR was derived either by using
our own data (GRB 080123) or by a re-analysis of archival data (GRB
150424A) using the equations relating \HII\ luminosity and SFR
published in \citet{Kennicutt1998} and \citet{Savaglio2009}. 
Additional details on individual targets are provided in the Appendix.
If not otherwise stated, SFR measurements based on
emission-line spectroscopy were not corrected for host-galaxy
extinction, slit-losses were usually taken into account.
   
\section{Results}
\subsection{Radio detection of the host of GRB 100206A}

With our deep radio observations we detected only one target in our
sample: the suspected host of GRB 100206A
(Table~\ref{Tab:targetlist}).   Using the \texttt{imfit} task under
CASA, at 5.5~GHz we measure a flux density of $F_\nu = 65 \pm 11
\mu$Jy, centered at R.A., Decl. (J2000) = 03:08:39.148 $\pm$ 0.017~s,
13:09:29.28 $\pm$ 0.15~arcsec. The flux we measure from this source is
in agreement with its reported non-detection by \cite{Berger2013.765},
who found $F_\nu$(5.8~GHz)$\lesssim80\mu$Jy ($5\sigma$).

\begin{figure}
\includegraphics[width=8.6cm,angle=0]{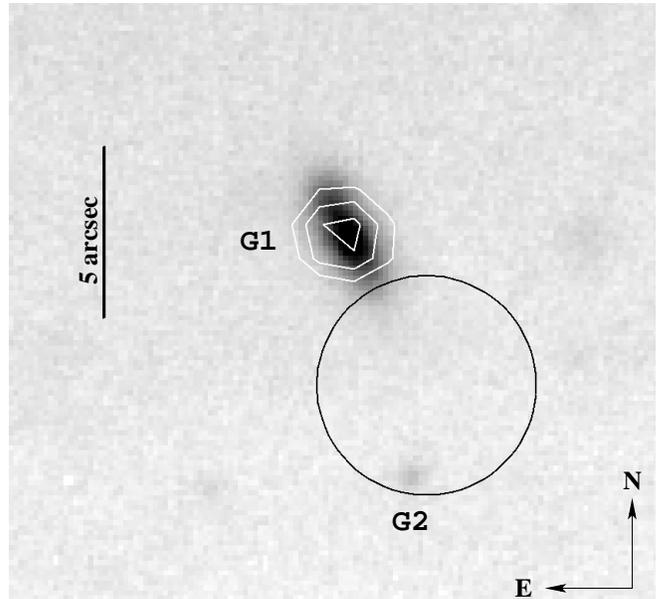}
\caption{ Radio 5.5\,GHz flux contours superimposed on a
  Gemini-N/GMOS $i$-band  image of the field of GRB 100206A (programme ID
  GN-2010A-Q-7, PI: D. Perley). The XRT error circle
  (90\% c.l.; see Appendix) is shown
  (radius 3.2 arcsec; as of Jan 2019), together with the suspected host
  galaxy (G1), and a fainter galaxy (G2) inside the error circle. Radio contour
  lines shown in white correspond to 3, 3.5, and 4 times the
  local 1$\sigma_{\rm rms}$ (Table~\ref{Tab:obs.summary}). At the redshift
  of GRB 100206A, 1 arcsec corresponds to 5.6 kpc projected distance.}
\label{fig:100206}
\end{figure}

The radio centroid (positional error  $\sim$0.2 arcsec) lies
5~arcsec away from the center of the  \swift/XRT 90\% c.l. error
circle, which at $z$=0.4068 corresponds to a  projected spatial
distance of $\sim$30~kpc. It coincides with the bulge of the
suspected GRB host galaxy (G1; Fig.~\ref{fig:100206}). Assuming a
Gaussian distribution for XRT position measurements, this angular
distance excludes the origin of the X-rays from the position of the
radio centroid with 99\% probability.  Conversely, with high
confidence we do not detect (Table~\ref{Tab:obs.summary}) radio
emission from sources inside the XRT error circle.  Therefore, it is
unlikely that the measured radio flux  is physically related to the
burst (e.g., the radio afterglow).

Furthermore, according to  \cite{Perley2012} the observed
[N\,{\sc II}]/H$\alpha$ and [O\,{\sc III}]/H$\beta$ emission line
ratios of this galaxy favor the interpretation that the strong
H$\alpha$ line is due to star formation and not associated with AGN
activity.  This would be consistent with evidence from the radio map;
using the \texttt{imfit} task under CASA, the radio source is
marginally resolved in one direction which would be unexpected in the
case of an AGN.

In order to calculate the radio SFR, we follow \cite{Greiner2016A&A593A}
\citep[with reference to][]{Murphy2011}, according to which\footnote{For a
recent discussion on the radio-based SFR calibration we refer to the
comprehensive analysis by \citet{Tabatabaei2017}, \citet{Mahajan2019}, and
\citet{Tisanic2019}.}  
\begin{equation}
\frac{{\rm SFR}_{\rm Radio}}{{\rm M}_{\odot}/{\rm yr}} = 
0.059\, \left({ F_\nu \over \mu {\rm Jy} }\right)\,
(1 + z)^{(\beta - 1)}\, \left(d_L \over {\rm Gpc} \right)^{2} 
\, \left( \nu \over {\rm GHz} \right)^{\beta}\,.
\label{Eq.JG}
\end{equation}
Here $d_L$ is the luminosity distance, $\beta$ is the spectral slope (unlike
those authors, we use the convention $F_\nu \sim \nu^{-\beta}$). 

If the observed radio flux is completely due to star-forming activity,
then it corresponds  to a SFR of $59\pm10$\,\msunyr (assuming
$\beta=0.7$). This is roughly twice the value derived by
\citet{Perley2012} from a study of the broad-band SED of this galaxy
in the wavelength range  between  $\sim$0.5 and 20$\mu$m (observer
frame).

This picture does not change when we perform a GRASIL analysis of the  SED of
this galaxy from the optical to the radio  band\footnote{for details of the
code and the set of templates see
\citet{Silva1998,Iglesias2007,Michalowski2010}},  analogous to what we have
done in \citet{Nic2014ApJ}. Adding now the radio detection to the data implies
a SFR of 63\,\msunyr, an IR luminosity of $\log (L_{\rm IR}/L_\odot)$=11.58, a
mass in stars of $\log (M_\star/M_\odot)$=11.42, and a global host visual
extinction of $A_V$=2.10 mag.

Recently, \cite{Perley2017} reported results of a
re-observation of several long-GRB host galaxies where substantial
radio-derived star formation rates in excess of 100\,\msunyr\ had
been found in previous studies. Surprisingly, for none of their 5 hosts
these authors could reproduce the previously reported large SFRs.
As stressed by \cite{Perley2017}, radio afterglow contamination,
noise fluctuations, and numerical artifacts produced by the image
processing could have affected some of the previous radio
detections.  In the case of GRB 100206A, our detection and 
radio flux measurement
is supported by three independent studies: The flux density we
measure at 5.5~GHz fits well with what has been predicted for this
galaxy based on its observed optical/infrared broad-band SED
(\citealt{Perley2012}, \citealt[][their figure 4]{Nic2014ApJ},
\citealt[][her figure 3]{Contini2018}).

Even though the host of GRB 100206A is forming stars at a high rate,
a physical link between the formation of the GRB
progenitor to the present epoch of star formation cannot be established.
Though the (90\% c.l.) XRT error circle touches the outer
regions of the suspected host galaxy (G1, Fig.~\ref{fig:100206}),
the large (projected) spatial offset between the burst position and
the centroid of the detected radio source disfavors
such a conclusion. 

\subsection{Non-detections and upper limits on star formation rates}

\label{Sec:RadioSFRs}
  
For the remaining 15 targets, we can only provide deep upper limits on the
SFR, as given in Table~\ref{Tab:SFRs}. In the case of our ATCA observations
the SFR is best constrained at 5.5~GHz. In the case of VLA all limits on the
SFR refer to observations in the C band (5.5~GHz). 

\setlength\tabcolsep{8.5pt}
\begin{deluxetable}{lllll}
\tablecolumns{7} 
\tablecaption{Radio-derived star-formation rates.}
\tablehead{
\colhead{GRB}  &
\colhead{$z$}  &
\colhead{SFR(radio) }&
\colhead{SFR(opt.)}&
\colhead{Ref.} \\
\colhead{} &
\colhead{} &
\colhead{\msunyr} &
\colhead{\msunyr} &
\colhead{} }
\startdata
{\rm ATCA}   &        &      &            & \\[1mm] 
050724  & 0.258   &$<$   8    & $<$0.05--0.17    & [1] \\ 
061006  & 0.438   &$<$  42    & 0.02       & [2] \\ 
061201  & 0.111   &$<$   1.5  & 0.14       & [3] \\    
070729  & 0.8     &$<$ 160    &            &     \\[1mm]
070809  & 0.219   &$<$   7    & 0.15       & [4] \\  
080123  & 0.496   &$<$  46    & 0.7        & [5] \\
130515A & 0.5     &$<$  45    &            &     \\ 
150424A & 0.298   &$<$  16    & 0.5--7     & [5] \\[3mm] 
{\rm VLA}    &        &        &            &  \\[1mm]
060502B & 0.287   &$<$  12    & 0.4--0.8   & [6] \\ 
061210  & 0.410   &$<$  25    & 1.2        & [6] \\ 
070724A & 0.457   &$<$  45    & 8.4 (+0.6,$-$6.1) & [11] \\
090621B & 0.5     &$<$  42    &            &     \\[1mm] 
100206A & 0.407   &$59\pm10$  & 30         & [8] \\  
100816A & 0.805   &$<$ 127    & $58^{+51}_{-26}$ & [7] \\
101224A & 0.5     &$<$  47    &            &     \\ 
130603B & 0.356   &$<$  22    & 1.7--5        & [6,9]  
\enddata 
\tablecomments{Radio-derived upper limits refer to 5$\sigma_{\rm rms}$ 
at 5.5~GHz (Eq.~\ref{Eq.JG}). In three cases no redshift is known and we
assumed $z$=0.5; for GRB 070729 we used a photometric redshift estimate from
\cite{Leibler2010} (see Appendix). If not otherwise stated, optical data
refer to emission-line measurements. See the Appendix for more details.  {\it
Additional notes to individual sources:}\  GRB 061201: the data refer to the
host-galaxy candidate G1.  GRB 070809: the data refer to the host-galaxy
candidate G1.  The host-galaxy candidate  G2 is not included in this table as
it is an elliptical galaxy.  GRB 100206A: the data refer to the host-galaxy
candidate G1.  GRB 130515A: the SFR is based on a \texttt{Le Phare} fit of the
broad-band SED obtained with GROND (Fig.~\ref{fig:130515}).  GRB 150424A: the
data refer to the bright spiral next to the OT.  The SFR based on optical
spectroscopy is 0.15\,\msunyr, while the SFR based on a \texttt{Le Phare} fit
of the galaxy's SED is $\sim$7\,\msunyr\ (Fig.~\ref{fig:150424sed}).  {\it
References:}\   
[1] \citet{Berger2005Natur,Prochaska2006ApJ,Malesani2007_050724},  
[2] \citet{Berger2009ApJ}, [3] \citet{Stratta2007}, [4] \citet{PerleyGCN7889},
[5] this work, [6] \citet{Berger2014ARAA}, [7] \citet{Kruhler2015}, [8]
\citet{Perley2012}, [9] \citet{Cucchiara2013ApJ,Fong2014ApJ...780,Ugarte2014}, 
[10] \citet{Kocevski2010MNRAS.404}, [11] \citet{Chrimes2018}. }
\label{Tab:SFRs}
\end{deluxetable}

\begin{figure}[t!]
\includegraphics[width=8.6cm]{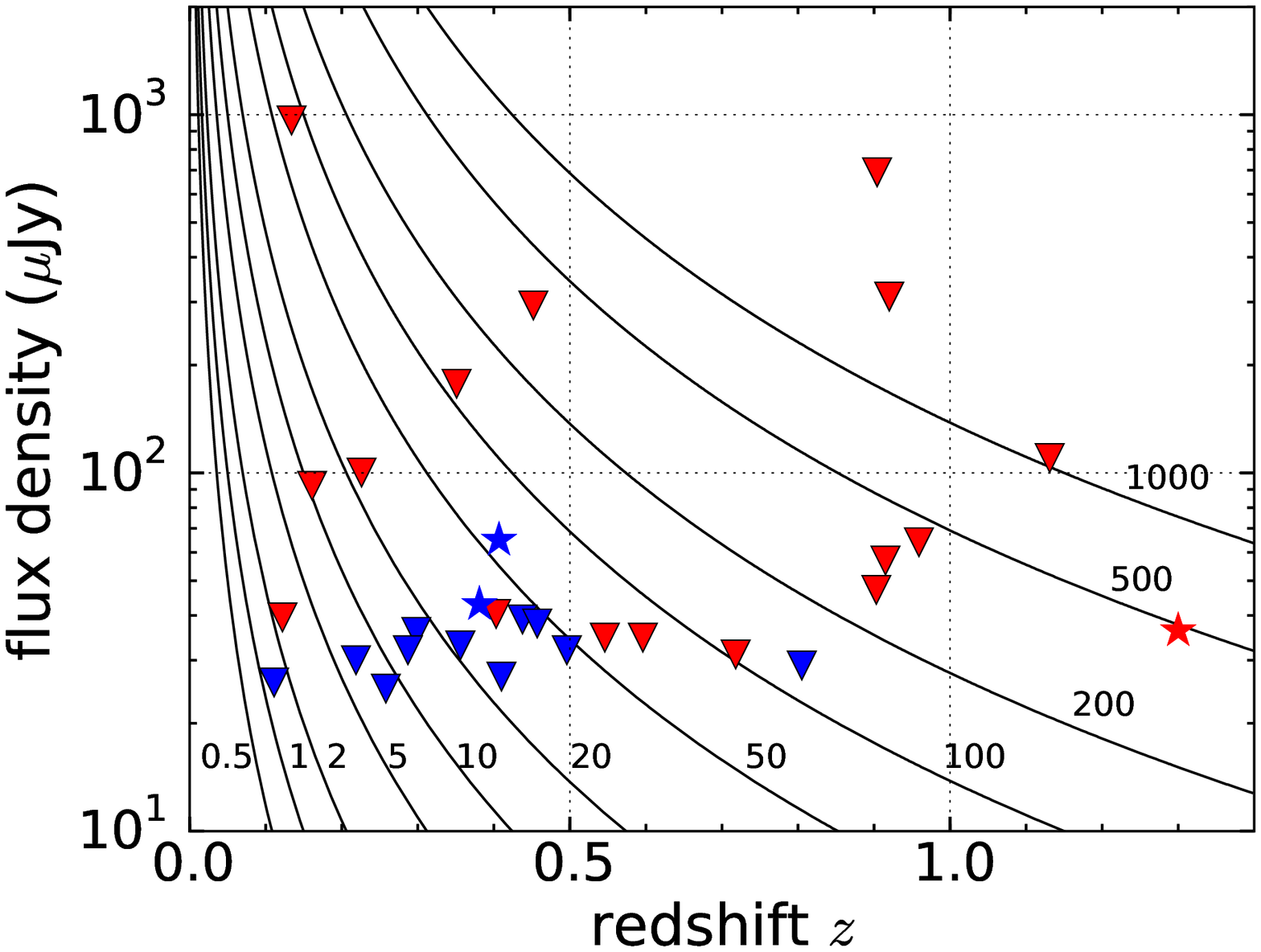}
\caption{The radio-continuum flux at 5.5~GHz from short-GRB host galaxies and
  the corresponding radio-derived star-formation rates
  (Eq.~\ref{Eq.JG}). Color coding: blue = this work
  (Table~\ref{Tab:SFRs}) as well as GRB 071227 \citep{Nic2014ApJ}, red = data from
  \cite{Fong2015ApJ815102} and \cite{Fong2016ApJ}. Detections are shown as
  stars (GRB 071227, 100206A, and 120804A), 5$\sigma_{\rm rms}$ upper limits
  as downwards pointing triangles. Solid  lines correspond to different
  star-formation rates (in units of \msunyr).}
\label{SFRSummary} 
\end{figure}

Figure~\ref{SFRSummary} summarizes the radio-derived  SFRs\footnote{A
similar plot is shown in \cite{Stanway2014MNRAS} but for long-GRB
host galaxies. Since these authors used a slightly different
equation for the transformation of radio flux into SFR, their curves
SFR($z, F_\nu$) do not perfectly match the curves shown
here. Because of analogous reasons, the radio-SFR calculated for GRB
120804A differs slightly from the SFR calculated by \citet[][their
equation 4]{Berger2013.765}.}.  For those cases in our target list
where more than one host-galaxy candidate is implicated, we only plot
the result for the host-galaxy candidate with the smallest redshift
(in no cases is this an elliptical galaxy). The four bursts from our
sample  without spectroscopic redshifts (GRB 070729, 090621B, 101224A,
and 130515A) are not plotted.  These data are supplemented by data
taken from \citet[][GRB 071227]{Nic2014ApJ}, and \citet[][GRB
120804A]{Berger2013.765}.  In addition, we  display data from
\citet[][their table 8]{Fong2015ApJ815102} and \citet[][their table
1]{Fong2016ApJ}.  All upper limits reported by these authors have
been transformed to an observing frequency of 5.5~GHz, assuming a
spectral slope $\beta$=0.7. In those cases where several observations
with different upper limits were reported for the same host, only the
most constraining data point is plotted.\footnote{Seven bursts in our
target list are also included in the summarizing table in
\citet{Fong2015ApJ815102}. Our radio upper limits are significantly
deeper, however. The field of GRB 050724, 070724A, and 130603B was
also observed with the VLA by \cite{Fong2016ApJ}; their upper limits
agree with ours.}

Figure~\ref{SFRSummary} shows that at 5.5~GHz a sensitivity of
better than about  40$\mu$Jy ($5\sigma$ rms) has been reached for
more than 50\% of the targeted galaxies. Reaching much deeper sensitivity
limits would require much more telescope time. 

If we consider a SFR sensitivity of 20\,\msunyr as a minimum we would
like to achieve for a target, then we are restricted to the lowest
redshifts, $z\lesssim0.3$. In our sample all 5 targeted
galaxies in this redshift range remained undetected. For the nearest
events at $z<0.2$ the inferred SFR upper limits do even reach 1.5 to
5\,\msunyr.  Though the sample size is small, one important result can
be stressed: In no case evidence for  intense star-forming activity
close to a GRB explosion site has been found.

Going to higher redshifts increases the sample size, but already for
targets with $z\gtrsim$0.5 the achieved SFR sensitivity is just
sufficient to potentially find the most extreme star-forming hosts: At
redshifts $z\gtrsim$0.5 even star-forming galaxies with a SFR of
$\sim$50\,\msunyr will not be detected anymore. All this naturally
limits the meaningfulness of such radio surveys.

Nevertheless, the data clearly show that the majority of short-GRB hosts is not
forming stars at a high rate, though some clearly do.
This raises the question whether the
percentage of heavily star-forming hosts
among the short-GRB host-galaxy ensemble is unexpectedly high.

\subsection{The most actively star-forming short-GRB hosts}

Among the 25 short-GRB host galaxies listed in \cite{Berger2014ARAA} which
have a measured SFR (bursts from mid 2005 to 2010), 18 hosts have  an optical
SFR $\lesssim2.5\,$\msunyr\ (in his list GRB 071227 has 0.6\,\msunyr). The
remaining seven hosts have an  optical SFR between 2.5 and 30\,\msunyr. Only
the host of GRB 100206A (in his list  30\,\msunyr) and GRB 101219A
(16\,\msunyr) lie above 10\,\msunyr. 

Following GRB 071227 (\citealt{Nic2014ApJ}), the host of GRB 100206A is
the second short-GRB host in our sample with a radio detection due to star-formation
activity. Together with the hosts of GRB 101219A
($z$=0.718; \citealt{Chornock11518}), GRB 100816A \citep[see
Table~\ref{Tab:SFRs};][]{Kruhler2015}, and  GRB 120804A \citep[SFR
$\sim$300\,\msunyr, $z\sim$1.3;][]{Berger2013.765},  these five hosts
represent the current sample of very actively star-forming galaxies that
hosted a short GRB (SFR $>$10\,\msunyr).  Though the sample size of these very
actively star-forming galaxies is small, it represents at least $\sim$10\% of
the short-GRB host-galaxy population (including all morphological types of
hosts).

Given this percentage, is this pointing to a subpopulation of
young NS-NS merger systems which were born in recent starbursts?
When exploring the consequences of the discovery of the heavily
star-forming hosts of GRB 100206A and GRB 120804A,
\citet{Perley2012} and \citet{Berger2013.765} pointed out that in
the redshift range 0.5-−1.0, i.e., close to the redshift range where
massive star-forming short-GRB hosts have been found
($z\gtrsim0.4$), the comoving number density of massive, luminous
and ultra-luminous infrared galaxies accounts for 10-20\% of the
total comoving SFR density of the Universe (\citealt{Casey2014}), but
only a small fraction of comoving stellar mass density
(\citealt{Caputi2006}). In this respect, the fraction of heavily
star-forming short-GRB hosts found so far is not remarkably high.
Thus there is currently no evidence for a
subpopulation of short-lived (young) short-GRB progenitors which
were born in recent starbursts. Future multi-wavelength studies of
very nearby hosts are required to address this question with better statistics.

\section{Constraints on radio emission from late-time short-GRB
  radiation components}

Even though our radio data were taken with the goal of searching for optically
hidden star-forming activity, our non-detections are well suited to set deep
constraints on late-time radiation components of short GRBs. In the following,
we quantify the observational limits that can be placed on flux from late-time
radio afterglows and late-time kilonova radio flares.  

\subsection{Constraints on late-time radio afterglow flux}

In the time window we are studying here (months to years  after the
corresponding burst), no radio afterglow of a short GRB has ever been
detected. The short burst with the latest  successful radio
observation is GRB 140903A, whose radio afterglow was still found
$\sim$10 days post burst
\citep{Fong2015ApJ815102,Zhang2017ApJ...835...73Z}. For
comparison, several long-GRB radio afterglows have been successfully
observed at host-frame times $>$100 days \cite[][their figure 7]{Chandra2012}.
The most extreme example here is the long GRB
030329 whose radio afterglow was still detected several years after
the burst \citep{Horst2008A&A...480...35V,Peters2019ApJ...872...28P}.

\citet{Chandra2012} performed a detailed analysis of 14 years of radio follow-up
observations of GRB afterglows with the VLA, including 35 short bursts (with
2 detections). These data confirmed what is known from the optical/X-ray bands
\citep[e.g.,][]{Nysewander2009ApJ...701..824N,Kann2011, Nicuesa2012}:
short-GRB afterglows are intrinsically dim sources,  in each wavelength region
being on average at least one order of magnitude fainter than long
GRBs. According to figure 8 in \citet{Chandra2012},  in our time window we
would expect an  afterglow radio luminosity  at 8.5~GHz far below $10^{29}$
erg s$^{-1}$ Hz$^{-1}$. Changing the observed frequency to 5.5~GHz does not
substantially change this conclusion. 

In order to calculate the $k$-corrected  luminosities  we would infer from our
radio observations, we followed \cite{Chandra2012} (note our convention for
$F_\nu$; Sect.~\ref{Sect.Intro}):
\begin{equation}
L_\nu = 4\pi d_L^2 \,F_\nu\,(1+z)^{\beta - 1}\,.
\end{equation}  
Following these authors, in Table~\ref{Tab:KN} the luminosity $L_{\nu,1}$
assumes an optically thin, flat, post-jet-break radio afterglow with a
spectral slope $\beta=-1/3$ (i.e., a positive radio slope) while $L_{\nu,2}$
assumes a kilonova radio flare with $\beta$ = 0.7 (see
Sect.~\ref{Sec:flares}). 

\setlength\tabcolsep{3.4pt}
\begin{deluxetable}{ll ccrr rrr}
\tablecolumns{9} 
\tablecaption{Upper limits on the luminosities
of radio afterglows and late-time radio flares.}
\tablehead{
\colhead{GRB}  &
\colhead{$z$}   &
\colhead{$dt/(1+z)$} &
\colhead{$F_\nu^{\rm lim}$}&
\colhead{$L_{\nu, 1}$} &
\colhead{$L_{\nu, 2}$} &
\colhead{$\nu L_{\nu, 1}$}&
\colhead{$\nu L_{\nu, 2}$}\\
\colhead{(1)} &
\colhead{(2)} &
\colhead{(3)} &
\colhead{(4)} &
\colhead{(5)} &
\colhead{(6)} &
\colhead{(7)} &
\colhead{(8)} }
\startdata
{\rm ATCA}   &   &&&     &      &            & \\[1mm] 
 050724 & 0.258 &  8.149 & 25 &   4.0 &   5.0 &   2.2 &   2.8\\
 061006 & 0.438 &  4.721 & 39 &  17.7 &  25.8 &   9.7 &  14.2\\
 061201 & 0.111 &  5.979 & 26 &   0.8 &   0.8 &   0.4 &   0.5\\
 070729 & 0.8   &  4.531 & 37 &  52.8 &  96.8 &  29.0 &  53.3\\[1mm] 
 070809 & 0.219 &  4.810 & 30 &   3.4 &   4.2 &   1.9 &   2.3\\
 080123 & 0.496 &  3.678 & 32 &  18.6 &  28.1 &  10.2 &  15.5\\
 130515A& 0.5   &  1.632 & 31 &  18.3 &  27.8 &  10.0 &  15.3\\
 150424A& 0.298 &  0.118 & 68 &  14.3 &  18.7 &   7.9 &  10.3\\[3mm]
{\rm VLA}   &   &&&     &      &            & \\[1mm] 
 060502B& 0.287 &  5.995 & 44 &   8.7 &  11.2 &   2.6 &   3.4\\
        &       &  6.556 & 32 &   6.3 &   8.2 &   3.5 &   4.5\\
 061210A& 0.410 &  5.569 & 27 &  10.6 &  15.1 &   5.8 &   8.3\\
 070724 & 0.457 &  4.912 & 38 &  18.8 &  27.7 &  10.3 &  15.2\\
 090621B& 0.5   &  4.219 & 29 &  16.8 &  25.5 &   9.2 &  14.0\\[1mm] 
 100206A& 0.407 &  3.322 & 65 &  25.6 &  36.4 &  14.1 &  20.0\\
 100816A& 0.805 &  2.298 & 29 &  42.4 &  78.0 &  23.3 &  42.9\\
 101224A& 0.5   &  3.215 & 32 &  18.9 &  28.7 &  10.4 &  15.8\\
 130603B& 0.356 &  1.066 & 33 &   9.8 &  13.5 &   5.4 &   7.4
\enddata 
\tablecomments{Column \#2 provides the used redshift, column \#3 the time after
  the burst in the GRB host frame in years.  Column \#4 lists the
  $5\sigma$ observed  upper limits of the flux density (in $\mu$Jy).  The
  following two columns provide the specific luminosities in 
  $10^{28}$ erg s$^{-1}$ Hz$^{-1}$ (assuming isotropic emission). The two last
  columns contain $\nu L_\nu$ in $10^{38}$ erg s$^{-1}$.  $L_{\nu, 1}$
  assumes a radio afterglow, $L_{\nu, 2}$ a kilonova radio flare
  (see text).  All luminosities refer to 5.5~GHz. }
\label{Tab:KN}
\end{deluxetable}

Table~\ref{Tab:KN} presents the constraints we place on the  late-time radio
luminosity of the 16 short-GRB events studied here. In the case of multiple
5.5~GHz observations  of the same target (GRB 061201, 090621B, 101224A) the
calculated luminosities refer to the combined data set, averaged over the time
of the observing runs (which were close to each other).  This averaging was
not done for GRB 060502B, since here the observations included two different
frequencies; we used only the result for the 5.5~GHz observation in October
2014.  Finally, an averaging was not done for GRB 150424A either; here only
the data from the observing run in June 2015 were taken into account (see
Appendix). 

Table~\ref{Tab:KN} shows that our observations  achieved a sensitivity between
0.1 and 5$\,\times\,10^{29}$ erg s$^{-1}$ Hz$^{-1}$.  Assuming that after its
peak a radio afterglow flux fades according to $F_\nu\sim t^{-1}$
\citep{Chandra2012}, our non-detection of radio emission from short-GRB
explosion sites months to years after a burst is not unexpected
(see figure 8 in \citealt{Chandra2012}).

Finally, we note that in our time window at 5.5~GHz radio emission from the
reverse shock is not expected to be detectable \citep[][their figure
1]{Resmi2016ApJ...825...48R}. 

\subsection{Constraints on late-time kilonova radio flares}

\label{Sec:flares}

According to the picture emerging from GW170817 and GRB 170817A, short
bursts originating from compact binary mergers with a NS component should be
followed by kilonova light
\citep{Lattimer:1974a,Symbalisty:1982a,Li:1998bw}.  Indeed, at least 5 events
from our target list showed evidence for an additional radiation component in
their optical/NIR afterglows: GRB 130603B
\citep{Berger2013ApJ,Tanvir2013Natur}, GRB 070809
\citep{Jin2019arXiv190106269J},  and  GRB 050724, 061210, and 150424A
\citep{Rossi2019}.
  
Several authors have pointed out that the interaction of the dynamical
mass ejecta  with the circumburst medium could produce long-lasting radio
emission, a radio flare \citep{Nakar2011,Margalit2015,
Hotokezaka:2015eja,Horesh2016,Hotokezaka:2018gmo,Radice:2018pdn}.  Moreover,
if the merger is followed by the formation of a rapidly spinning magnetar
(period $P\sim$1~ms), a deposition of this energy into the ejecta could result
in an even brighter flare
\citep{Metzger2014MNRAS.437.1821M,Fong2016ApJ,Horesh2016,Kathirgamaraju2019}.

Given the expected peak time, light curve, and potentially high radio
luminosity, a kilonova radio flare  could be unambiguously identified on
times scales of 1 to 10 years \citep[e.g.,][]{Radice:2018pdn}. If
detected, radio flares could act as the smoking  gun of past kilonova
events. However, no such signal has yet been reported.

\begin{figure}[t!]
\includegraphics[width=8.6cm]{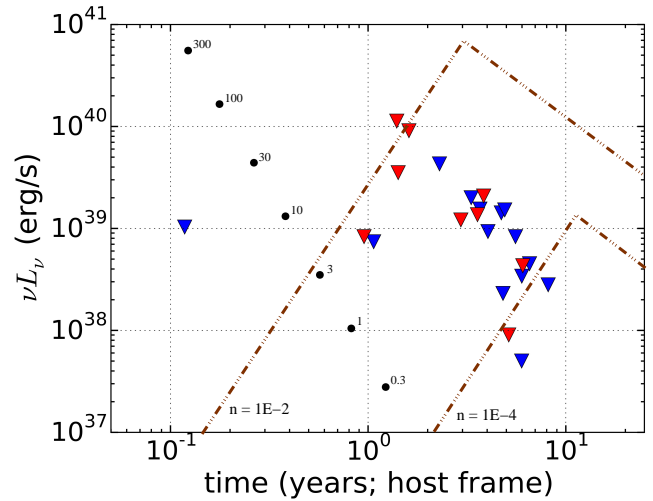}
\caption{Observational  constraints on the 5.5~GHz radio luminosity
  $\nu L_\nu$ (erg s$^{-1}$)  for 19 short GRBs, mostly years after
  the corresponding event.  In blue are shown our data
  (Table~\ref{Tab:KN}), in red data from \cite[][their table
    1]{Fong2016ApJ}, transformed to an observing frequency of
  5.5~GHz. Note that the results for GRB 050724, 070724A, and 130603B
  are included here twice since these events were independently
  observed by these authors (at slightly different times after the
  corresponding burst). In the case of GRB 060502B, the results for
  two visits are plotted (Table~\ref{Tab:obs.summary}).  For
  comparison, we show in dark filled points the predicted  peak
  luminosity and peak time for kilonova radio flares  inferred from
  the numerical merger simulations reported in
  \citet{Radice:2018pdn}. Here we used ejecta kinetic energy of
  $10^{50}$ and present  the results as a function of the circumburst
  gas density $n$ (in units if cm$^{-3}$). In addition, two models  of
  magnetar-powered radio flares from \citet[][their  figure
    1]{Fong2016ApJ} are also shown (for $n=10^{-2}$ and
  $10^{-4}$cm$^{-3}$; thick dashed-dotted lines).  For more details,
  see main text. }
\label{fig:knlight} 
\end{figure}

Figure~\ref{fig:knlight} summarizes the upper limits we can set for a
kilonova radio signal for a total of 19 events. These include:
\vspace{-0.5\baselineskip}
\begin{enumerate}[(i)]
\setlength{\itemsep}{0pt}
\setlength{\parskip}{0pt}
\item 12 short bursts with known redshift from the sample discussed here
      (Table~\ref{Tab:targetlist});
\item 6 additional short bursts  with
      known redshift studied by \cite{Fong2016ApJ}: GRB 051221A, 080905A,
      090510, 090515, 100117A, 101219A;
\item the short GRB 071227 \citep{Nic2014ApJ}. 
      Even though in this case the host was detected
      in the radio band,  the emission does not arise from the position of
      the optical afterglow but from the central bulge of the host
      ($z$=0.381). The radio data ($dt$(host frame)=4.038 yr) then constrain the
      flux from the GRB explosion site to 
      $F_\nu$(5.5~GHz) $<35~\mu$Jy (5$\sigma_{\rm rms}$), i.e., 
      $\nu L_\nu < 9.3\,\times\,10^{38}$ erg s$^{-1}$.  
\end{enumerate}
We note that for some events several host-galaxy candidates
exist. Analogous to Table~\ref{Tab:SFRs}, in these cases we used the
redshift of the cosmologically less distant galaxy.  
Not included in Fig.~\ref{fig:knlight} are those 4 events in our sample where
no spectroscopic redshift is known (GRB 090621B, 070729, 101224A, 130515A).

For comparison, in Fig.~\ref{fig:knlight} we also plot the peak
luminosity $\nu L_\nu$ computed at 5.5~GHz and the peak time emission
of the radio flare produced by the interaction of the dynamical ejecta
with the surrounding interstellar medium as given by the analytical
estimate of \citet{Hotokezaka:2015eja}. The model assumes ejecta
kinetic energy $E_{\rm kin} = 10^{50}$ erg and velocities $v_{\rm ej} =
0.3$~c as inferred from numerical relativity simulations
\citet{Hotokezaka:2018gmo,Radice:2018pdn}. The microphysical
parameters are fixed to $\epsilon_B$= 0.01, $\epsilon_e$=0.1, $p$=2.4.
We also plot a model for a magnetar-powered radio flare as it is
discussed by \cite{Fong2016ApJ} and where we have chosen two values
for the circumburst gas density ($n$= 10$^{-2}, 10^{-4}$ cm$^{-3}$), a
rotational energy of the magnetar of  $E_{\rm rot} = 10^{53}$ erg and
an ejected mass of $M_{\rm ej}$ = 0.03\,\msun (adopting $\epsilon_B$=
0.1, $\epsilon_e$=0.1, $p$=2.4).

Our negative observational result augments previous unsuccessful searches for
late-time radio flares following short GRBs
\citep{Metzger2014MNRAS.437.1821M,Fong2016ApJ,Horesh2016}. Even though most of
our data explore a parameter space similar to what was already discussed by
these authors, we provide four more events with secure host-galaxy
identification and spectroscopic redshift (GRB 061006, 061210A, 100206A,
100816A). GRB 100206A is included in this list,  since the radio emission does
not arise from a position inside the \swift/XRT error circle but from the
central bulge of the suspected host (Fig.~\ref{fig:100206}). In other words,
this is not kilonova emission.

The observations imply that in the time window between about 1 and 10 years
(host frame) kilonova radio flares exceeding a luminosity of $\sim 10^{39}$
erg s$^{-1}$ are either rare or do not exist. In principle, a general
non-detection of these flares can still be explained within the allowed broad
parameter space of the underlying models, including the not well constrained
microphysical parameters $\epsilon_B$ and $\epsilon_e$ and the
uncertainties on the circumburst number density $n$. In any case, 
the potentially very luminous radio flares predicted by 
models which rely on magnetar-powered energy deposition into the ejecta
were not found in our investigation.

\section{Summary and conclusions}

We performed a deep radio-continuum survey of 16 short GRB hosts, with the
original goal to search for  optically obscured  star-forming activity.  Only
one host was detected (GRB 100206A; $z$=0.407). Its detection was not
surprising, however, since a high radio flux was already expected based on its
observed broad-band SED in the optical/IR bands
\citep{Perley2012,Nic2014ApJ}. 

Combining our radio data with published data shows that at least 5 of
about 40 short-GRB hosts compiled by \cite{Berger2014ARAA} have a relatively
high SFR of $>$10\,\msunyr\ (GRB 071227, 100206A, 100816A, 101219A,
120804A). 

The present data base show that galaxies forming stars at a high rate
are not uncommon among the short-GRB host-galaxy population.  However,
whether or not  this implies  a physical link between recent
star-forming activity and the formation of short-lived short-GRB
progenitors is not clear. At least in the case of GRB 071227 and
100206A, the radio emission  which traces the star-formation activity
does not arise from the GRB explosion site. Furthermore, at least 3 of
these 5$-$6 hosts are characterized by a relatively large stellar mass
\citep[GRB 071227, 100206A, 120804A; see][their table
1]{Nic2014ApJ}. As noted by several authors (e.g.,
\citealt{Leibler2010,Fong2013,Berger2014ARAA}), it could just be the
large reservoir of stars which increases the probability of a NS-NS
merger to occur (for detailed stellar population synthesis models
see, e.g., \citealt{Tauris2017ApJ...846..170T,Belczynski2018}.)

In conclusion, in the local universe ($z<$0.3) star-forming short-GRB host
galaxies are not forming stars at very high rates. There is no observational
evidence for optically obscured star formation. The first extensively
star-forming  short-GRB hosts appear in the redshift range from $z$= 0.3 to
0.5 (GRB 071229 and 100206A). The data suggest that beyond $z$=0.3 a
non-negligible fraction of short-GRB hosts is forming stars at a high rate. 

In addition to our search for optically obscured star-formation
activity, we used our radio data to derive deep upper limits on the
luminosity of the corresponding GRB afterglows and predicted late-time
radio flares following short-GRB kilonovae. While the general
non-detection  of a radio afterglow on timescales of months to years
after a short burst is not surprising, our data provide valuable
constraints on kilonova magnetar models.  By adding 12 more events
with redshift information, we substantially increased the available
data set for  quantifying the luminosities of radio flares in the time
frame between 1 and 10 years (host-frame time) post burst.  Their
general non-detection places severe constraints on the most
extreme kilonova models (e.g., \citealt{Fong2016ApJ}). The existence of
very bright and long-lasting radio flares as predicted by
these models is not supported by the data.

In order to draw more definitive conclusions on the age distribution of the
short-GRB progenitor population and the existence or non-existence of radio
flares, more and better data are needed.  In particular, hosts at low
redshifts may act as future ``Rosetta Stone events'';  the host of GRB 170817A
at $d\sim$40~Mpc  \citep{Abbott2017ApJ...848L..12A,Coulter2017Sci...358.1556C,
Smartt2017Natur.551...75S,Kim2017} is the first such promising case in this new
sample of LIGO/Virgo-selected events. The future discovery of  gravitational
wave signals from short-GRBs in cosmologically local hosts suggests
that the age distribution of short-GRB progenitors and the brightness of kilonova 
radio flares  will be estabished more firmly in the near future.

\begin{acknowledgements}

S.K. and A.N.G. acknowledge support by the Th\"uringer Ministerium
f\"ur Bildung, Wissenschaft und Kultur under FKZ 12010-514 and  by
grants DFG Kl 766/16-1 and 766/16-3. L.K.H. is grateful for funding
from the INAF PRIN-SKA 2017 program 1.05.01.88.04.
M.J.M.~acknowledges the support of the National Science Centre, Poland
through the SONATA BIS grant 2018/30/E/ST9/00208 and the POLONEZ grant
2015/19/P/ST9/04010; this project has received funding from the
European Union's Horizon 2020 research and innovation programme under
the Marie Sk{\l}odowska-Curie grant agreement
No. 665778. A.R. acknowledges support from Premiale LBT 2013.
S.B. acknowledges support by the EU H2020 under ERC Starting Grant,
no.~BinGraSp-714626.
This paper is based on observations collected with the Australia
Telescope Compact Array (ATCA; programme ID C2840, PI: A. Nicuesa  Guelbenzu),
the NSF's {\it Karl G. Jansky} Very Large Array (VLA; programme ID 13B-313,
14B-201, 15B-214, PI: A. Nicuesa Guelbenzu), VLT (ESO programme ID 088.D-0657,
PI: A. Nicuesa Guelbenzu and programme ID 383.A-0399, PI: S. Klose), as well
as GROND at the MPG 2.2m telescope at ESO La Silla Observatory (PI:
J. Greiner), BUSCA at the Calar Alto 2.2m (PI: J. Gorosabel) and publicly
available data obtained from the ESO-VLT, the Gemini and the Hubble Space
Telescope data archives.
The Australia Telescope is funded by the Commonwealth of Australia for
operation as a National Facility managed by CSIRO.  S.K. and A.N.G. thank
Catarina Ubach \& Sarah Maddison, Swinburne University, Ivy Wong,
Martin Bell, \& Mark Wieringa CSIRO
Sydney, and Jamie Stevens, CSIRO Narrabri, for very helpful discussions and
observing guidance. In particular, Sarah Maddison
substantially helped performing remote observations.
The National Radio Astronomy Observatory is a facility of the National 
Science Foundation operated under cooperative agreement by 
Associated Universities, Inc. 
S.K. and A.N.G. thank the staff at NRAO (Socorro, NM)
for their hospitality during two visits,
in particular Dale Frail, Heidi Medlin, Drew Medlin, and J\"urgen Ott.
We thank the staff at NRAO for performing the observations.
This publication makes use of data products from the Wide-field Infrared
Survey Explorer, which is a joint project of the University of California, Los
Angeles, and the Jet Propulsion Laboratory/California Institute of Technology,
funded by the National Aeronautics and Space Administration.
This work made use of data supplied by the UK Swift Science Data Centre at the
University of Leicester.
Part of the funding for GROND (both hardware
and personnel) was generously granted by the Leibniz-Prize to G. Hasinger (DFG
grant HA 1850/28-1) and by the Th\"uringer Landessternwarte Tautenburg. 
We thank the referee for a very careful reading of the manuscript
and many valuable suggestions which helped to improve the paper
substantially.
\end{acknowledgements}



\begin{appendix}

\section{Notes on individual targets}

{\it General:}\ Enhanced \swift/XRT error circle data were taken from
http://www.swift.ac.uk/xrt$_-$ positions/ \citep{Goad2007,Evans2009}.  In
the following all reported XRT coordinates with their corresponding error
radii refer to January 2019 (90\% c.l.).

Follow-up observations of short-GRBs in the radio band are summarized in
\citet{Chandra2012} and \citet{Fong2015ApJ815102}.  The following bursts from
our target list have reported  radio observations within some days after the
event:  GRB 050724, 061210A, 070724, 070729, 090612B, 101224A, 130603B, and
150424A.  Detected were only the radio afterglows of GRB 050724, 130603B, and
150424A \citep[see][their table 8]{Fong2015ApJ815102}. 


{\it GRB 050724}:\ The initial pulse of the burst was hard and had a FRED-like
profile. It was followed by softer emission; $T_{90}$(15-350 keV)= $3 \pm 1$~s
\citep{Krimm3667}. Its observed spectral lag is consistent with those of short
bursts \citep{Gehrels2006}.   The afterglow and its host galaxy are in detail
discussed in the literature
\citep{Berger2005Natur,Gorosabel2006A&A,Prochaska2006ApJ,
Malesani2007_050724,Fong2010}.
The  optical afterglow was located in the outskirts of a relatively bright
galaxy ($R\sim$20 mag) at a redshift of $z = 0.258$. A global host extinction
of $A_v^{\rm host}\sim$0.4 mag was derived by \cite{Gorosabel2006A&A}. The
radio afterglow was detected with the VLA 0.6 days after the burst at 8.46~GHz
($F_\nu$ = 173~$\mu$Jy).  It was only seen again one day later at 8.46~GHz
\citep[$F_\nu$ = 465~$\mu$Jy;][]{Berger2005Natur,Fong2015ApJ815102}.

Several authors have classified the host as an elliptical galaxy
\citep[e.g.,][]{Berger2005Natur}. Though \citep{Malesani2007_050724} found
that it shows a faint outer spiral structure, suggesting that is a disk galaxy
with a light-dominating central bulge. This conclusion was strengthened by
\HST\ images \citep[see figure 3 in][]{Fong2010}.

Based on optical spectroscopy, the galaxy has a very low star formation
rate. Reported upper limits are 0.02\,\msunyr\
\citep{Berger2005Natur}, 0.05\,\msunyr\
\citep{Prochaska2006ApJ}, and 0.17\,\msunyr\
\citep{Malesani2007_050724}. 


{\it GRB 060502B}:\ The \swift/BAT lightcurve consists of two spikes.
According to  \citet{Sato5064}, $T_{90}$(15-350~keV) = 90$\pm$20~ms, while
according to \citet{Krimm5704,Avanzo2014MNRAS} it is $T_{90}$(15-350~keV) =
140~ms. No optical afterglow was found. The burst and its suspected host are
in detail discussed by \cite{Bloom2007ApJ...654..878B}. According to these
authors,  the most likely host-galaxy candidate is a relatively bright
early-type galaxy galaxy ($R\sim$ 18.7 mag, $z$=0.287) $\sim$17.5 arcsec south
of the XRT error circle (see their figure 1). However, publicly available
Gemini-N/GMOS images (programme ID GN-2006A-Q-14 PI: B. Schmidt) taken in May
2006 clearly reveal a disk component with at least one spiral arm in both
directions. Based on the \OII~ emission line, \cite{Bloom2007ApJ...654..878B}
calculated a SFR of 0.4--0.8~M$_\odot$ yr$^{-1}$.  \cite{BergerApJ664} (their
figure 1) pointed out that  a much fainter galaxy ($R\sim25.8$ mag)  is
located inside the XRT error circle. Its redshift is not known.


{\it GRB 061006}:\ The burst began with an intense double-spike, followed by
lower-level persistent emission. Including   this radiation component the
duration is  $T_{90}$(15-350~keV)=130$\pm$10~s \citep{Krimm5704}.  The
burst was also seen by the Suzaku Wideband All-sky Monitor (WAM) and
$T_{90}$(50~keV -5~MeV)=0.42~s \citep{Urata5717}.  The optical afterglow
was found with VLT/FORS1 \citep{MalesaniGCN5705,MalesaniGCN5718}.  In
Gemini-S/GMOS images the host galaxy appears as a faint ($R_C\sim$24.0 mag)
and fuzzy object \citep[see figure 1 in][]{BergerApJ664}.  Its shows weak
emission lines from the \OIII~ doublet, H$\beta$, and \OII~ at a common
redshift of $z=0.4377\pm0.0002$. The \OII~ line flux corresponds to a SFR of
0.02\,\msunyr\ \citep{Berger2009ApJ}.

\HST\ observed the field in several occasions between  October 2006 and
November 2007 using ACS in the NIR (filter F814W; programme ID 10917,
PI: D. Fox). The host is an inclined, nearly edge-on spiral galaxy. The
optical afterglow was located close to the galactic bulge 
\citep[see figure 9 in][]{Fong2010}.


{\it GRB 061201}:\ The burst had a double-peak structure, there is no evidence
for extended emission. Its duration was  $T_{90}$(15-350 keV) = $0.8 \pm
0.1$~s \citep{Markwardt5882}; \citet{Avanzo2014MNRAS} give $T_{90}$(15-350
keV) = 0.78~s. On VLT/FORS2 images the  suspected host is a bulge-dominated
galaxy  $\sim$17 arcsec NW from the afterglow \citep[in the following G1; see
  figure 5 in][]{Stratta2007}. Long-slit spectroscopy  revealed emission lines
from \OII~ and H$\alpha$ at a common redshift of $z=0.111$. Similar results
were  reported based on Magellan observations \citep{BergerGCN5952}. The SFR
based on the H$\alpha$ flux is 0.14~M$_\odot$ yr$^{-1}$ \citep{Stratta2007}. 

The field was observed by \HST\ in three different epochs between 2007 and 2013
(programme ID 12502, PI: A. Fruchter).  Although the \HST\ image goes deeper
than the VLT image, at the position of the optical transient there is no
evidence for an underlying galaxy. However, inside the 1\farcs4 XRT error
circle lies a faint galaxy (G2) that is not visible on the VLT/FORS2
images. It is  located $\sim$1\farcs8 south-east of the optical transient
\citep[$r\sim$25.5 mag; see figure 10 in][]{Fong2010}. There is no redshift
information. The position of this galaxy close to the optical transient
classifies  it as second host-galaxy candidate.


{\it GRB 061210}:\ The \swift/BAT lightcurve shows a hard FRED-like spike,
followed by extended, soft emission;  $T_{90}$(15-350 keV) is $85 \pm 5$~s
\citep{Palmer5905}.  The burst was also seen by Suzaku/WAM and $T_{90}$(50
keV-5 MeV) = 0.047~s \citep{Urata5917}. 

No optical afterglow was found.  A host-galaxy candidate was first identified
by \citet{Berger5922}, its redshift is 0.41 \citep{Cenko5946,BergerApJ664}.
On archived Gemini-N/GMOS images the host appears as a bulge-dominated galaxy
\citep[see figure 1 in][]{BergerApJ664}.  According to \citet{Berger2014ARAA},
its SFR is 1.2\,\msunyr.


{\it GRB 070724A}: \ In the \swift/BAT energy band the burst showed a single
spike. According to \citet{Parsons6656}, $T_{90}$(15-350~keV) was
$0.4\pm0.04$~s, while according to \citet{Avanzo2014MNRAS} it was 0.43~s. The
discovery of the optical/NIR afterglow was reported by
\cite{Berger2009ApJ...704..877B} (see their figure 1). These authors also
studied the host galaxy which lies at a redshift of $z$=0.457  \citep[see
also][their figure 3]{Kocevski2010MNRAS.404}.  The optical/NIR afterglow was
affected by about 2 mag host-galaxy visual extinction along the line of sight
\citep{Berger2009ApJ...704..877B}, indicating that this is a dust-rich
host. Optical spectra revealed a global SFR between about 1 and 100~M$_\odot$
yr$^{-1}$; internal host-galaxy extinction made a more robust measurement
difficult \citep{Kocevski2010MNRAS.404}. According to \cite{Berger2014ARAA},
the SFR is 2.5\,\msunyr.

\begin{figure}
\includegraphics[width=8.8cm,angle=0]{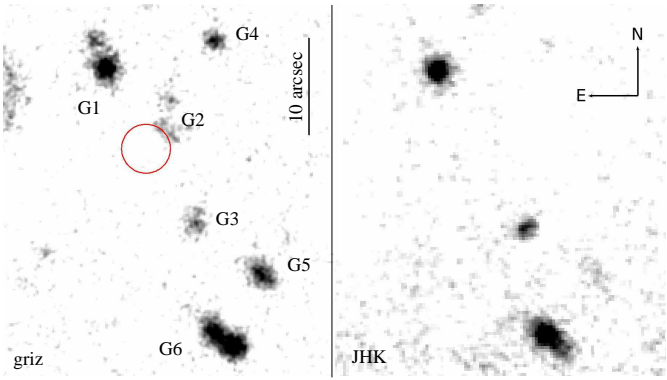}
\caption{GROND combined $g'r'i'z'$ and $JHK_s$-band image of the field of GRB
  070729. Also shown is the \swift/XRT  error circle ($r$=2.5
  arcsec). Galaxies  close to the error circle are labeled. No radio source
  has been found in this field. (The extended faint  structure seen in the
  $g'r'i'z'$-band image close to the upper left corner  is a ghost image from
  a bright star.)}
\label{fig:070729}
\end{figure}

{\it GRB 070729}:\ The \swift/BAT light curve showed two overlapping peaks.
According to \citet{Sato6681}, $T_{90}$(15-350 keV) = $0.9 \pm 0.1$~s, while
\citet{Avanzo2014MNRAS} measure $T_{90}$(15-350 keV) = 0.99~s.  Follow-up
observations did not reveal an afterglow, either in the optical \cite[for
GROND  see][]{Aybueke2008AIPC.1000} or in the radio band
\citep{Chandra6742}. At the western border of the enhanced XRT error circle
lie two (or three) faint objects \citep[G2;][]{Aybueke2008AIPC.1000}. 
Several additional brighter galaxies lie some arcsec
away (G1, G3, and G4-G6; Fig.~\ref{fig:070729}). 

The field was also studied by \cite{Leibler2010}. Based on their broad-band
photometry these authors derived a photometric redshift for G3 of
$z=0.8\pm0.1$. No spectroscopic redshifts are known. The GROND data suggest
that G1, G3, and G6 are dusty star-forming galaxies, probably at the same
redshift. G1 and G6 have been detected by the \WISE satellite (G1 is listed in
Table~\ref{tab:short_WISE}). G1 and G3 are very red objects $((r'-K_s)_{\rm
  AB} \sim$ 3~mag). We consider G1-G3 as host-galaxy candidates.  None of
these labeled galaxies (G1-G6) has been detected in our ATCA observing run.


{\it GRB 070809}:\ The \swift/BAT light curve shows a single peak.  According
to \citet{Krimm6732}, $T_{90}$(15-350 keV) = $1.3 \pm 0.1$~s, while according
to \citet{Avanzo2014MNRAS}, it is $T_{90}$(15-350 keV) = 1.28~s.  The recent
possible detection of kilonova light  in the GRB afterglow confirms the
short-burst nature of this event \citep{Jin2019arXiv190106269J}. 
 
Observations with the Keck telescope did not reveal an underlying host galaxy
either at the position of the optical afterglow or within the 3\farcs6 XRT
error circle \citep{PerleyGCN7889}. The galaxy nearest to the optical
afterglow is an edge-on spiral 6 arcsec north-west  \citep[named ``S2'' in
figure 2 in][]{Berger2010}. At a redshift of $z=0.2187$ this galaxy lies at
a projected distance of $\sim$22 kpc from the optical afterglow.  Its flux  in
the \OII~ line corresponds to an optical SFR of $0.15~$M$_\odot$ yr$^{-1}$.  A
second galaxy \cite[named ``S3'' in][]{Berger2010} lies 6 arcsec west from the
optical transient.  It is an E0 elliptical at a  redshift of $z=0.473$, with
no evidence for star formation.

The field was also observed by \HST\ in Aug 2009 and May 2010 with two
different filters (F606W/F160W; programme ID 12502, PI: A. Fruchter). At the
position of the optical afterglow there is no underlying host galaxy either at
the optical ($>$28.1 mag) or in the NIR \citep[$H>$26.2 mag;][]{Fong2013}. The
\HST\ observations  revealed the presence of a faint background galaxy inside
the XRT error circle \citep[named ``S1'' in][]{Berger2010}. Its redshift is
not known. 

No radio flux was detected either from the three designated galaxies or 
from the position of the optical transient.

\begin{figure}
\includegraphics[width=8.8cm,angle=0]{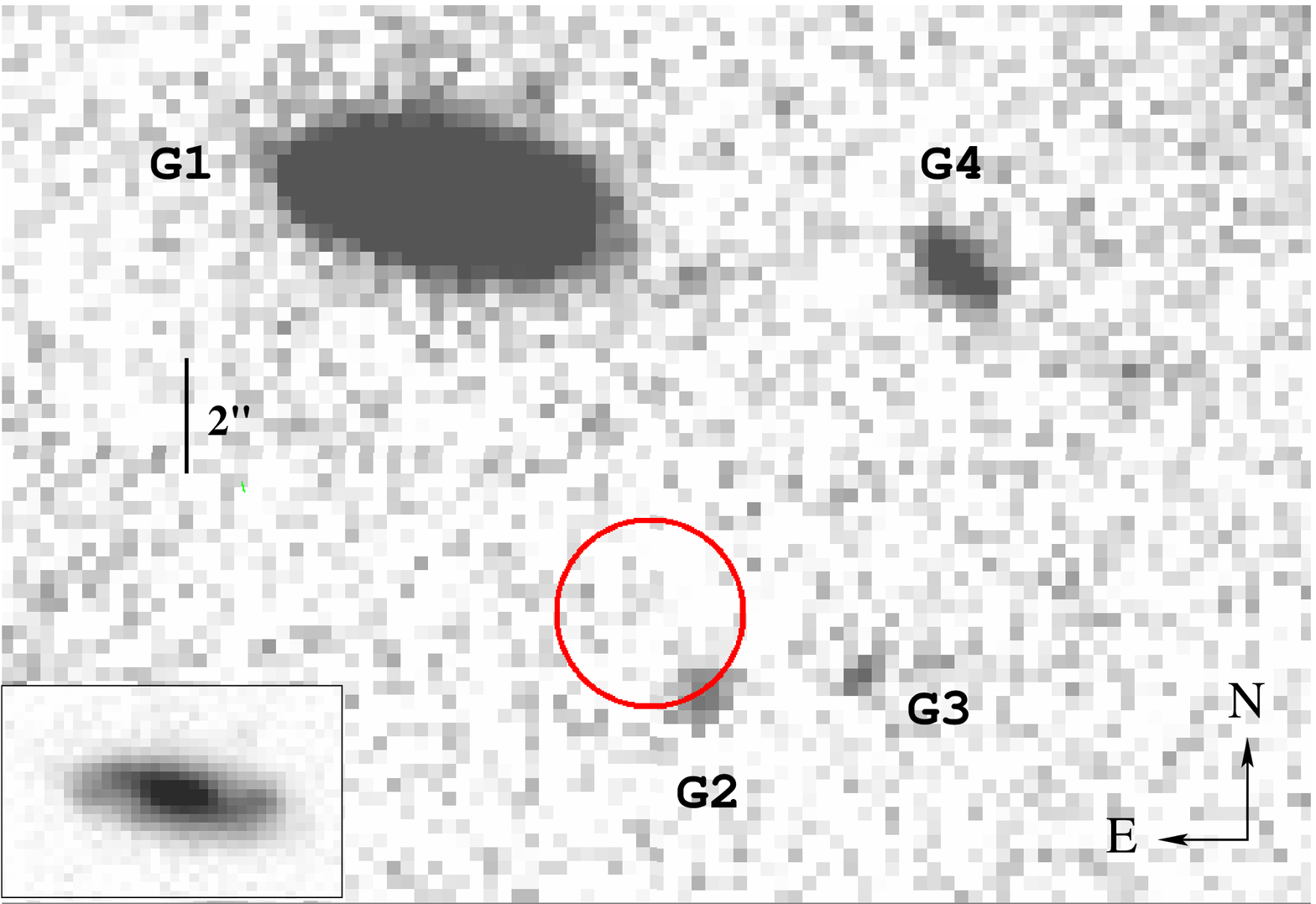}
\includegraphics[width=8.8cm,angle=0]{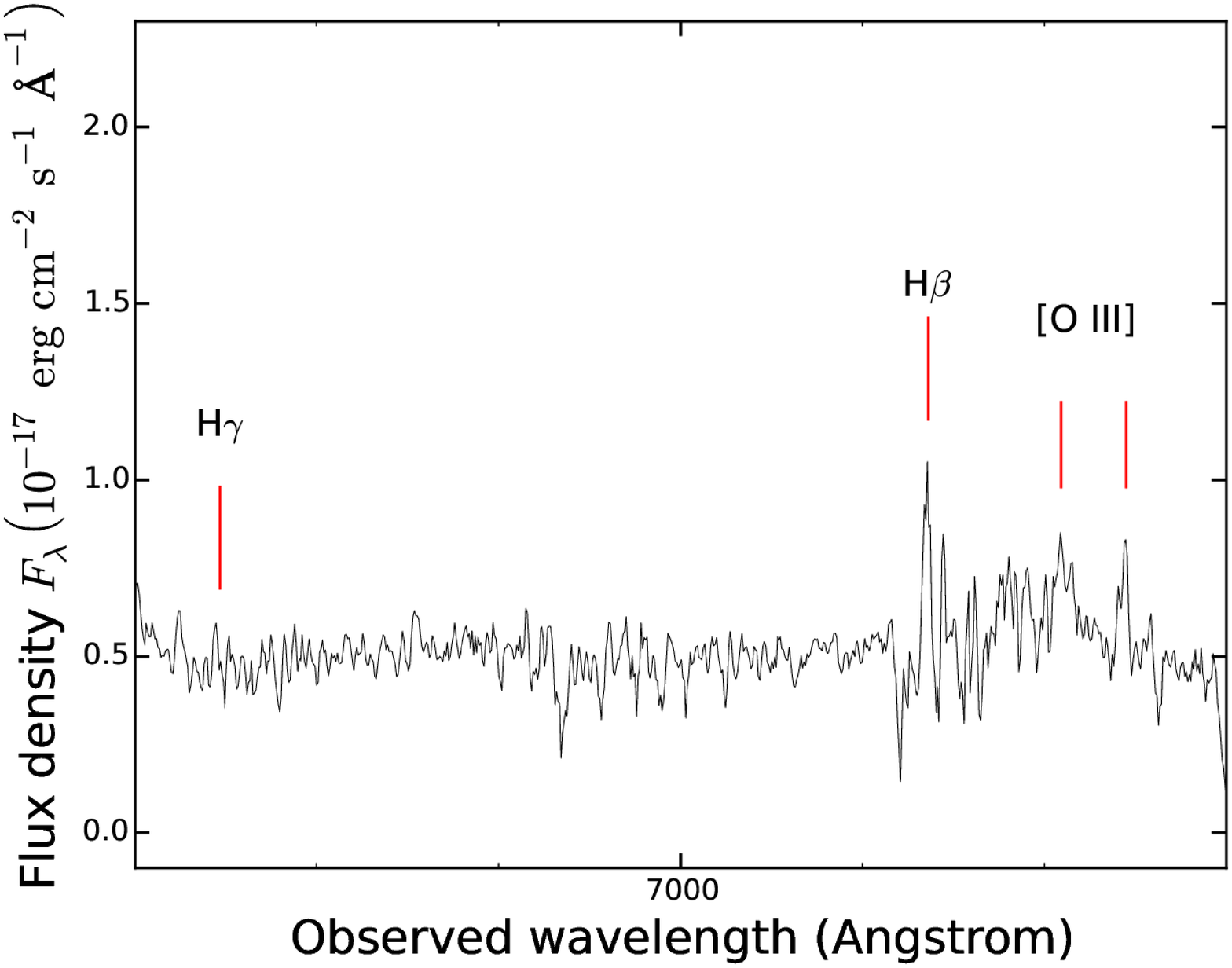} 
\caption{{\it Top:}\ VLT/FORS2 $R_C$-band image of the field of GRB 080123. In
  red it is shown the $r$=1\farcs7 XRT error circle.  Among the four galaxies
  close to the error circle G1 is the most likely host-galaxy candidate
  \citep{Leibler2010}. The inset shows the spiral structure of G1 as
  it appears after changing the contrast parameters in the VLT image.  {\it
    Bottom:}\ VLT/FORS2 spectrum of G1.  It shows emission from the
  \OIII~ doublet, H$\beta$ and H$\gamma$ at a common redshift of $z$=0.496.}
\label{fig:080123} 
\end{figure}

{\it GRB 080123}:\ In the \swift/BAT energy window  the burst showed a
FRED-like profile, followed by  soft, extended emission. $T_{90}$(15-350 keV)
was $115 \pm 30$~s \citep{Tueller7205}. This was refinded to  115.18~s by
\citet{Avanzo2014MNRAS}.  Suzaku/WAM also detected the burst with a duration
$T_{90}$(50 keV-5 MeV) of about 0.40 seconds \citep{Uehara7223}pNo optical
afterglow was found.  VLT/FORS2 $R_C$-band and ISAAC $K_s$-band imaging of the
field was obtained in May 2009 (ESO programme ID 383.A-0399; PI:
S. Klose). Inside the 1\farcs7 XRT error circle, the VLT images do not reveal
any source down to $R_C=27.0$ mag and $K_s=23.3$ mag.  An optical faint source
(G2) lies at the border of the XRT error circle, another source around 3
arcsec away (G3; Fig.~\ref{fig:080123}). G2 and G3 are very faint  in the
$R_c$ band ($\sim$25.5 mag)  and not detected  down to $K_s=23.3$ mag.

The most prominent object in the field is a galaxy (G1, $z$=0.496;
$R_C$=19.6 mag) $\sim$8 arcsec north-east of the center of the XRT error
circle \citep{Leibler2010}. It is the most likely host-galaxy candidate.
The FORS2 $R_C$-band image reveals that this is an inclined, barred spiral
with two symmetric spiral arms.

Long-slit spectroscopy of G1 was obtained  with VLT/FORS2 in October and
November 2011 using the {\it GRIS 600B} and {\it GRIS 600RI} grism (PI:
A. Nicuesa Guelbenzu, ESO programme ID 088.D-0657), covering the wavelength
range from 3300~\AA~ to 8450~\AA. The spectral slit was oriented along the
major axis of the galaxy G1. The spectrum shows  emission lines of H$\beta$,
H$\gamma$, and \OIII~ at a common redshift of $z$=0.496, in agreement with
what was reported by \cite{Leibler2010}. The blue spectrum obtained with the
{\it GRIS 600B} is of less quality; no emission or absorption lines could be
identified.  The (not extinction-corrected) SFR based on the measured H$\beta$
line flux in the FORS2 spectrum is $0.7~$M$_\odot$ yr$^{-1}$. 

No redshift could be derived for G2. Although during the FORS2
observations the spectral slit was  covering this galaxy, no trace of this
object could be identified.  None of these galaxies was detected in our radio
observations.

{\it GRB 090621B}: \ The bursts showed a single peak, in the \swift/BAT energy
band $T_{90}$(15-350 keV) = $0.14 \pm 0.04$~s \citep{Krimm9551}. The burst was
also detected by Fermi/GBM and $T_{90}$(8-1000 keV) = 0.128~s
\citep{Goldstein9562}. The X-ray afterglow was rather faint
\citep{Beardmore2009GCN..9550}, no optical afterglow was found.  A single,
faint object inside the XRT error circle  \citep[$I \sim 22.9$
mag;][]{Levan2009GCN..9547,Galeev2009GCN..9549,Cenko2009GCN..9557} turned
out to be an M star \citep{Berger2009GCN..9559}. No host-galaxy candidate has
been reported so far. 

\begin{figure}
\includegraphics[width=8.8cm]{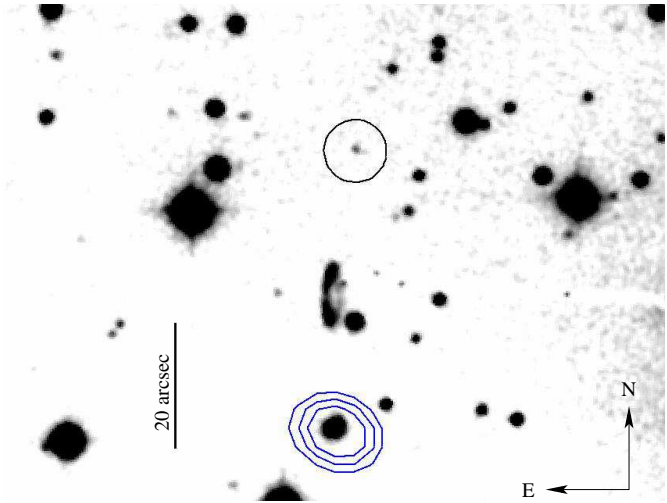}
\caption{$i$-band image of the field of GRB 090621B taken with BUSCA at the
  Calar Alto 2.2m telescope in July 2012.
  According to \citep{Berger2009GCN..9559}, the faint source inside the XRT
  error circle (in black; radius $r$=3.3 arcsec) is an  M star.
  Overlaid are the VLA radio contours in
  the field at 5.5~GHz (in blue; corresponding to 3, 4, and 5~$\mu$Jy
  beam$^{-1}$.).  A bright radio source lies south of the XRT error circle.
  Note that the elongated object between the XRT error circle and the radio
  source is an artifact in the image.}
\label{fig:090612}
\end{figure}

Our VLA radio data reveal a radio source ($F_\nu$(5.5 GHz)= 58$\pm 6$
$\mu$Jy), 45 arcsec south of the XRT error circle), at coordinates R.A.,
Decl. (J2000) = 20:53:53.70, 69:00:55.7. A comparison with our optical
observations of the field in July 2012 using the Calar Alto 2.2m telescope
(observer: J. Gorosabel) shows at this position a galaxy
(Fig.~\ref{fig:090612}).  Its morphology cannot be determined
with certainty. On our 2.2m/BUSCA $i$-band 
image its size is about $4\farcs5\,\times\,5\farcs2$, with the
long axis oriented in northwest-southeast direction. Its redshift is
not known. The GRB progenitor exploded at a projected distance of
$\sim$280/340/360~kpc (for $z$=0.5/0.8/1.0) from this object.

{\it GRB 100206A}: \ In the \swift/BAT energy window the burst showed a
single spike, $T_{90}$(15-350 keV) = 0.12$\pm$0.03~s
\citep{Sakamoto10379}. Also Fermi/GBM detected the burst; $T_{90}$(8-1000
keV) = $0.13 \pm 0.05$~s  \citep{Kienlin10381}.  No optical afterglow was
found. \cite{Perley2012} performed a comprehensive study of the burst and its
host and concluded that this is dusty, infrared bright galaxy ($L_{\rm IR}
\sim 4\,\times\, 10^{11}\,$L$_\odot$) at $z$=0.4068 with a global host
extinction of $A_{\rm V}^{\rm host}$ = 2 mag and a star formation rate of
about 30~M$_\odot$ yr$^{-1}$. According to its position in the \WISE
color-color diagram \citep{Wright2010}, this is a starburst galaxy
(Sect.~\ref{Sect:Targets}).  \cite{Perley2012} have pointed out that close to
this galaxy (in the following G1) lies a $\sim$4 mag fainter galaxy (in the
following G2). Its redshift is 0.803. Several arguments suggest that G1 is the
primary host galaxy candidate while G2 is a background  galaxy
(Fig.~\ref{fig:100206}).\footnote{Note that the  revised XRT coordinates,
R.A., Decl. (J2000) = 03:08:39.00, 13:09:24.8 (radius 3.2 arcsec; as of Jan
2019) are slightly different from the XRT coordinates used in
\cite{Perley2012}.} 


{\it GRB 100816A}: \ The burst was detected by \swift/BAT and Fermi/GBM.  In
the BAT energy window it consisted of a single spike with a duration
$T_{90}$(15-350 keV) = 2.9$\pm$0.6~s \citep{Markwardt2010GCN.11111,
Avanzo2014MNRAS}, while for GBM $T_{90}$(50-300 keV)$\sim$2~s
\citep{Fitzpatrick11124}.  The nature of this burst is debated. With its
duration the burst lies intermediate  between long and short-population
bursts. Its spectrum was hard, characteristic for a short burst, but with a
low peak energy \citep{Fitzpatrick11124}.    Its observed spectral lag in the
\swift/BAT energy window is consistent with zero, though the errors are
relatively large \citep{Norris2010,Bernardini2015}.  Other burst  and
afterglow properties move the burst closer to the long-burst population
\citep{Avanzo2014MNRAS}.

Its optical  afterglow was found by \swift/UVOT \citep{Oates11102}. It  was
well offset from its host \citep{Im11108}; its redshift is $z=0.8049$
\citep{Tanvir11123,Gorosabel11125}. No  late-time supernova signal was
detected down to 0.1 the peak luminosity of SN 1998bw
\citep{Tunnicliffe2012IAUS}, supporting the interpretation that GRB 100816A
belongs to the short-burst population.

\begin{figure}[t]
\includegraphics[width=8.8cm,angle=0]{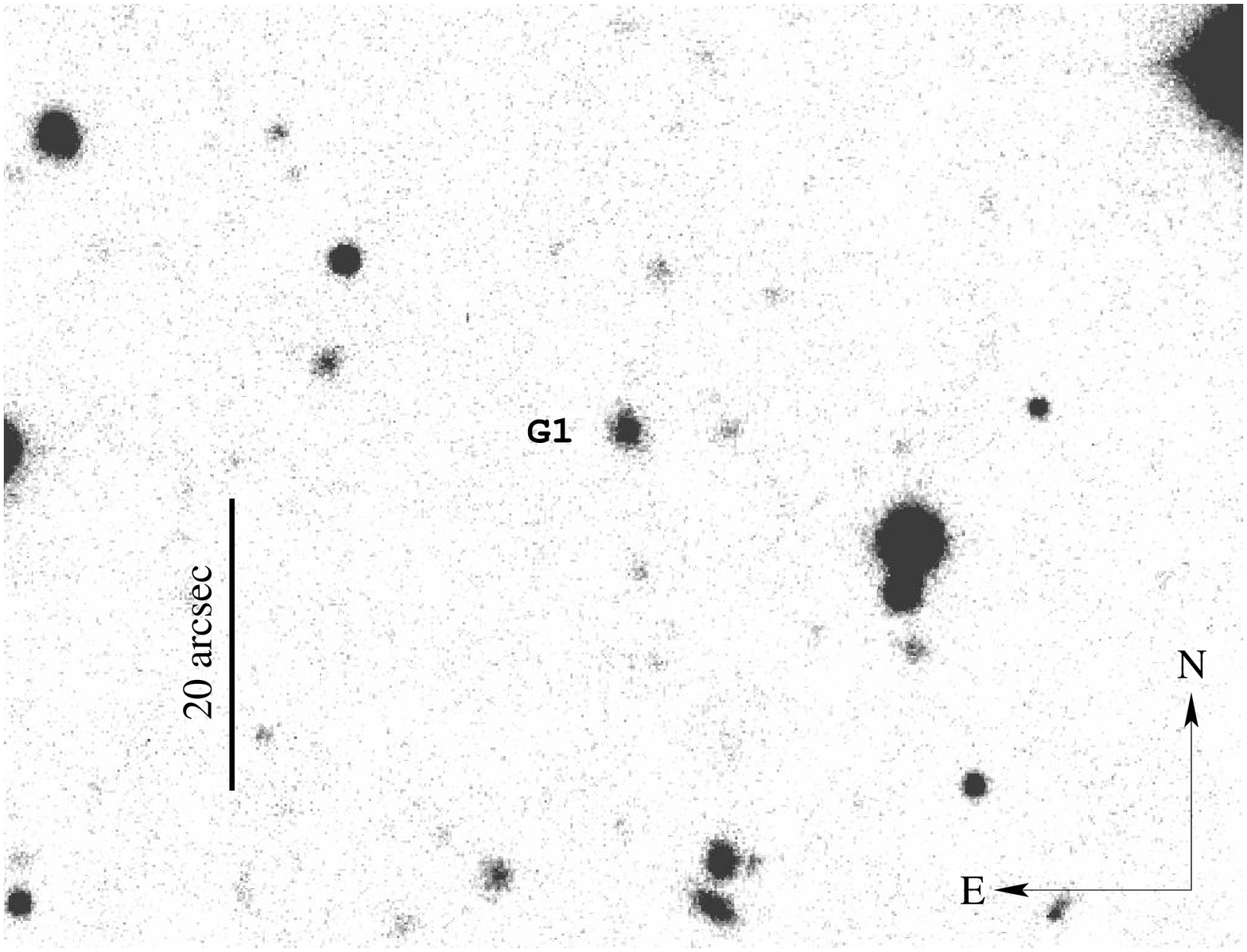}
\includegraphics[width=8.8cm,angle=0]{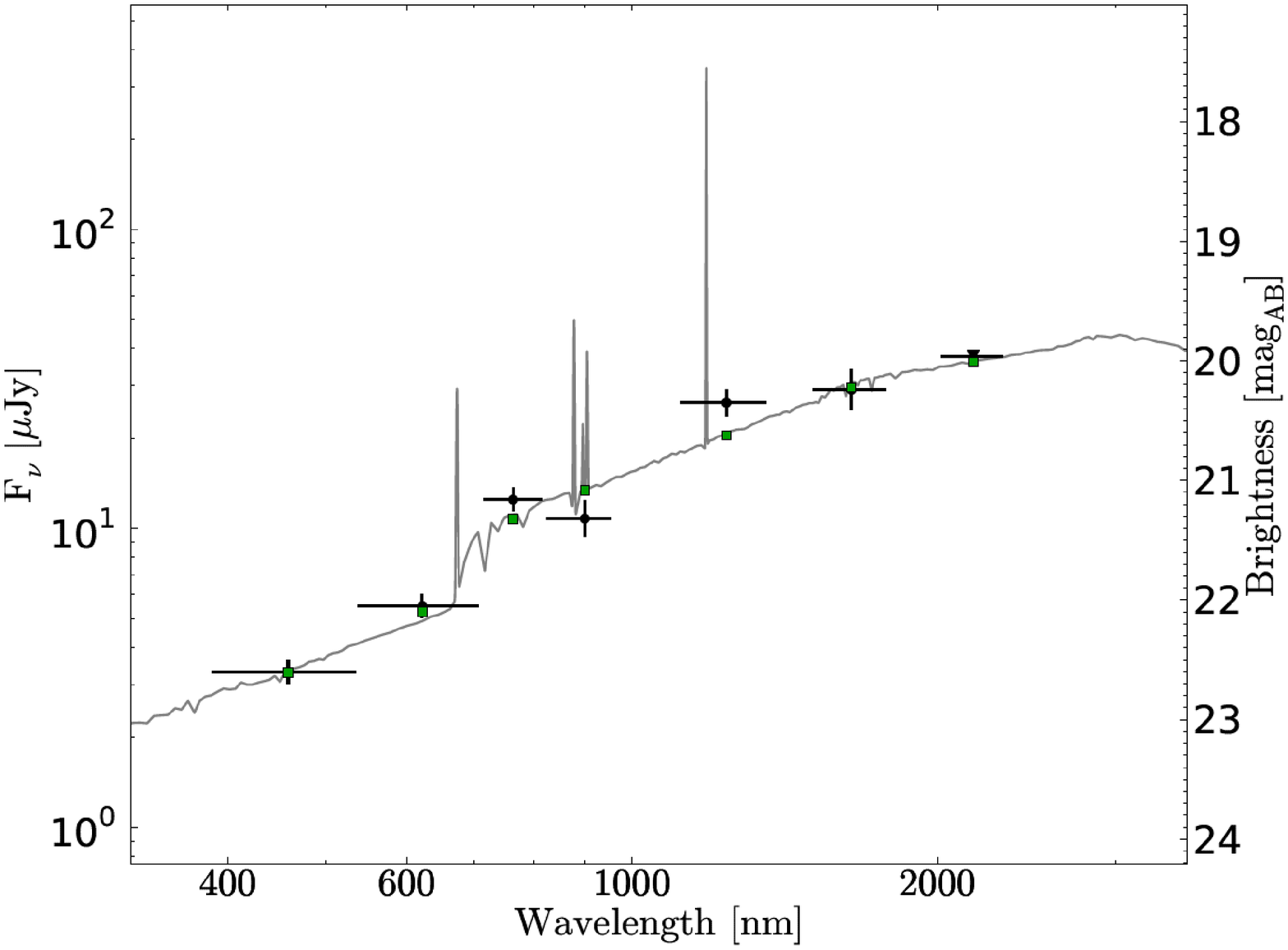}
\caption{{\it Left:} \ GROND $g'r'i'z'$ (white)-band image of the host galaxy
  of GRB 100816A (G1) taken on 18 August 2010. There is no OT detected
  anymore. There is no radio source in this field.  {\it Right:} \ Broad-band
  ($g'r'i'z'JHK_s$) SED of G1 based on GROND multi-color data and the best fit
  using the \texttt{Le Phare} package  \citep[corrected for a Galactic
  reddening along the line of sight of $E(B-V)=0.096$~mag;][]{Schlegel1998}.} 
\label{fig:100816}
\end{figure}

The afterglow was situated in the northern region of its host, about  1.3$''$
away from the host galaxy center (for the given redshift this corresponds to
about 10 kpc projected distance). In the combined GROND white-band the galaxy
appears morphologically disturbed in north-east direction. It is  a
bulge-dominated galaxy seen nearly face-on (G1; Fig.~\ref{fig:100816}). 
The host galaxy of GRB 100816A was also detected by \WISE in the W1
and W2 bands, but not in W3 and W4. 

Based on VLT/X-Shooter spectra, \citet{Kruhler2015} find  that the host is a
dusty galaxy ($E(B-V)_{\rm host}$ =  1.32 (+0.24, $-$0.22) mag) with an SFR of
$58^{+51}_{-26}$ M$_\odot$ yr$^{-1}$ and   a metallicity of
12+log(O/H)=8.75$^{+0.16}_{-0.18}$.  The observed mass-metallicity relation
for galaxies at $z\sim0.7$
\citep{Mannucci2009MNRAS.398.1915M,Mapelli2018MNRAS.481.5324M} then implies a
mass in stars of about log M/\msun\,$= 9.2 - 10.2$. The galaxy was also
studied by \citet{Perez2013EAS61} using the Spanish 10m Gran Telescopio
CANARIAS (GTC) and the telescopes at Calar Alto. These authors found that the
SED is best fit by a starburst galaxy. According to our GROND photometry, this
is a luminous (M$_B=-$21.7 mag), dusty ($E(B-V)_{\rm host}$ = 0.4 mag),
massive (log M/M$_\odot$ = 10.4$\pm$0.4) galaxy with an SFR  between 40 and
400~M$_\odot$ yr$^{-1}$ (Fig.~\ref{fig:100816}). 

\begin{figure}
\includegraphics[width=8.8cm,angle=0]{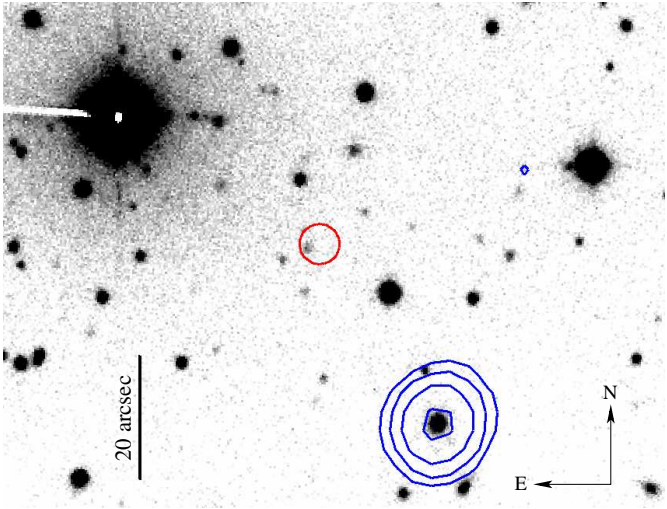}
\caption{Calar Alto 2.2m/BUSCA $i$-band image of the field of GRB 101224A,
  taken in June 2012 (observer: J. Gorosabel). In red is shown the $r$=3.2
  arcsec XRT error circle. Inside the error circle lies a faint galaxy
  \citep[$R\sim$21.5~mag;][]{Nugent2010GCN.11491,Xu2010GCN.11492}. In blue is
  shown a contour plot of our VLA 5.5~GHz observations. A bright radio source
  lies  south-west of the  center of the XRT error circle. Contour lines
  correspond to 2, 5, 10, 20~$\mu$Jy beam$^{-1}$.}
\label{fig:101224}
\end{figure}

{\it GRB 101224A:} \ In the \swift/BAT band the burst showed a multi-peak
structure, $T_{90}$(15-350 keV) = $0.2 \pm 0.01$~s
\citep{Markwardt11486,Avanzo2014MNRAS}.  It was followed by a rather faint
X-ray afterglow which was rapidly fading \citep{Paganini2010GCN.11488}. No
optical afterglow was detected
\citep{Kuroda2010GCN.11487,Landsman2010GCN.11490}, no radio afterglow was
found \citep{Zauderer2011GCN.11620}.  A faint object inside the XRT error
circle was discovered by \citet{Nugent2010GCN.11491}. It was later confirmed
by  \citet{Xu2010GCN.11492} and found to be a galaxy. Its 
morphology cannot be determined on our images (Fig.~\ref{fig:101224}).

Our VLA observations reveal a radio source ($F_\nu$(5.5~GHz)=
235$\pm$7~$\mu$Jy) 35 arcsec south-west of the XRT error circle at coordinates
R.A., Decl. (J2000) = 19:03:39.90, 45:42:20.5 (Fig.~\ref{fig:101224}).
Its redshift is not known. 

\begin{figure}[t]
\includegraphics[width=8.8cm,angle=0]{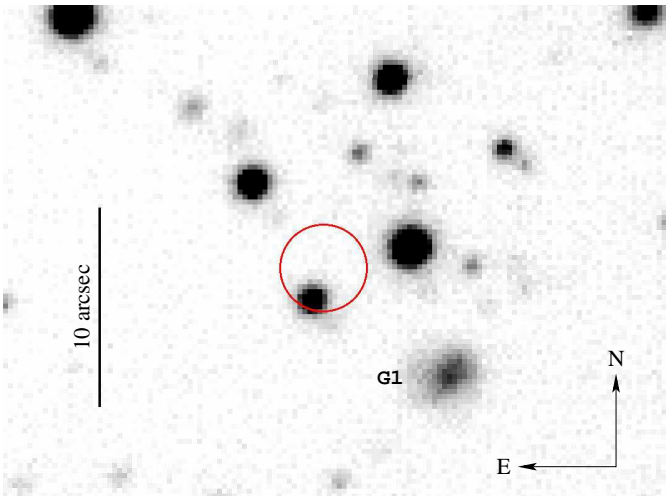}
\caption{VLT/FORS2 $R_C$-band image of the field of GRB 130515A (ESO programme
  ID 091.D-0558;  PI: A. Levan). The suspected host galaxy G1 lies $\sim$8.5
  arcsec south-west of the  center of the XRT error circle (radius 2.2
  arcsec).  There is no radio source in this field.}
\label{fig:130515}
\end{figure}

{\it GRB 130515A:}\ The burst was seen by several instruments: (i) \swift/BAT:
$T_{90}$(15-150~keV)   = 0.29$\pm$0.06~s \citep{Barthelmy14658,
Avanzo2014MNRAS}, (ii) Suzaku/WAM: $T_{90}$(50~keV-5~MeV) = 0.25~s
\citep{Iwakiri14688}, (iii) Fermi/GBM: $T_{90}$(50-300~keV) = 0.26~s
\citep{Jenke14663}.   No optical afterglow was found.  Using archived
VLT/FORS2 $R_C$-band data \citep[ESO programme ID 091.D-0558;  PI:
A. Levan;][]{Levan14667}, the galaxy closest to the 2\farcs2  XRT error
circle lies $\sim$8.5 arcsec south-west of the XRT position at coordinates RA,
Decl. (J2000) = 18:53:44.98, $-$54:16:50.9 (Fig.~\ref{fig:130515}). Its
spectroscopic redshift is not known. 

{\it GRB 130603B}: \ The burst was detected by \swift/BAT. It showed a single
FRED-like spike, $T_{90}$(15-350 keV) = $0.18 \pm 0.02$~s
\citep{Barthelmy14741,Avanzo2014MNRAS}.  Its observed spectral lag in the
\swift/BAT energy window was $-3.44^{7.27}_{5.58}$ ms, consistent with zero
\citep{Bernardini2015}.  The X-ray/optical  afterglow and the host galaxy has
been studied in detail by several groups
\citep{Cucchiara2013ApJ,Fong2014ApJ...780,Ugarte2014}. These authors found
that the afterglow was located  in a star-forming galaxy at $z$=0.3565.  The
galaxy is very dusty. The line-of-sight extinction to the optical afterglow
was about $A_V$(host)=0.9 mag, while the global visual extinction  of this
galaxy is about 1.3 mag. Its extinction-corrected SFR  is about 2 to 5
M$_\odot$ yr$^{-1}$.

The radio afterglow was detected with the VLA 0.37 days after the burst
at 4.9~GHz ($F_\nu$ = 125$~\mu$Jy) and 6.7~GHz ($F_\nu$ = 119$~\mu$Jy). It was 
only seen again the following day at 6.7~GHz
\citep[$F_\nu$ = 65$~\mu$Jy;][]{Fong2014ApJ...780,Fong2015ApJ815102}.

Our non-detection of this galaxy at 5.5 GHz is in agreement with
the predicted radio flux of the host in model 'MS5' discussed by
\citet[][her figure 4]{Contini2018}. According to this model
at 5.5~GHz one expects $\log \nu F_\nu \sim -18.2$, while we measure 
$\log \nu F_\nu < -17.8$ (in units of erg cm$^{-2}$ s$^{-1}$).


{\it GRB 150424A}:\ In the \swift/BAT energy window the burst showed a bright
multi-peaked episode followed by extended emission, $T_{90}$(15-350 keV) = $91
\pm 22$~s  \citep{Barthelmy17761}. The burst triggered also Konus-Wind. Here
it  showed a multipeak structure with a total duration of about 0.4~s
\citep{Golenetskii17752}. The optical/X-ray afterglow parameters  and its
physical implications are in detail discussed in
\cite{Kaplan2015ApJ...814L..25K}, \cite{Knust2017A&A...607A..84K}, and
\cite{Jin2018ApJ...857..128J}. The radio afterglow was discovered with VLA about
18~hr after the burst \citep[$F_\nu$ (9.8~GHz)
$\sim31~\mu$Jy;][]{Fong2015GCN.17804}.

\begin{figure}[t!]
\includegraphics[width=8.8cm,angle=0]{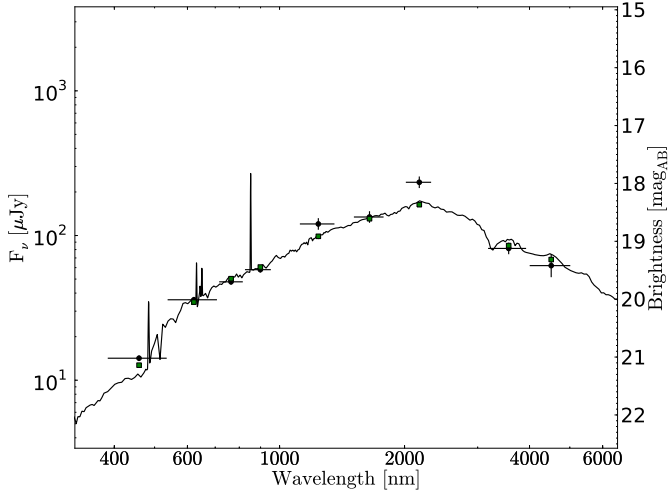}
\caption{GRB 150424A: \texttt{Le Phare} best fit of  the broad-band SED of
  the suspected host galaxy G1 based on GROND  $g'r'i'z'JHK_s$-band data as
  well as the publicly available \WISE satellite (W1, W2 filter)  data
  \citep[corrected for a Galactic reddening along the line of sight of
  $E(B-V)=0.06$~mag;][]{Schlegel1998}. }
\label{fig:150424sed}
\end{figure}

Follow-up observations performed with the Spanish GTC 10m telescope (programme
ID GTC72-15A; PI: A. Castro-Tirado) revealed the presence of a spiral  galaxy
2 arcsec  south-west to the optical transient \citep[G1; see figure 1
in][]{Jin2018ApJ...857..128J} and figure 2 in
\citet{Knust2017A&A...607A..84K}.  \citet{Castro-Tirado2015GCN17758} reported
a redshift of $z = 0.3$ for G1.  In a reanalysis of this spectrum we found
emission from H$\alpha$, \SII~ as well as the \NII~ doublet at a common
redshift of $z = 0.2981\pm0.0001$.  The flux in the H$\alpha$ line is  more
prominent in the western part than in the eastern part of the galaxy
(28\,$\,\times\, 10^{-17}$ erg s$^{-1}$ cm$^{-2}$ vs. 13\,$\times\, 10^{-17}$
erg s$^{-1}$ cm$^{-2}$). In total its flux corresponds to a SFR of
0.15\,\msunyr. Since the spectral slit covered only 1/3 of the apparent width
of the galaxy a more realistic estimate  of the optical SFR  is 0.5\,\msunyr.

Figure~\ref{fig:150424sed} shows the broad-band SED of G1. Fixing the
redshift to $z$=0.3, \texttt{Le Phare} finds that this is a star-forming
galaxy (SFR $\sim$7 M$_\odot$ yr$^{-1}$) with an internal reddening of
$E(B-V)_{\rm host}$= 0.25 mag. Its mass in stars is about 10$^{10}$ M$_\odot$.

Follow-up observations with HST  revealed the presence of a faint object
underlying the  position of the optical afterglow
\citep{Tanvir2015GCN18100}.  According to these authors this is a faint
galaxy at a redshift $z>0.7$. This is the second host-galaxy candidate (G2). 

The VLA detected the radio afterglow 18 hr after the burst at 9.8 GHz as a
faint source ($F_\nu$ = 31~$\mu$Jy) It was not detected anymore 4.8 and 7.9
days after the burst \citep{Fong2015GCN.17804}.

ATCA observations of the field  were performed in June and October 2015,
$dt_1$= 0.15 and $dt_2$= 0.50 yr after the burst. Unfortunately, during the
October run the source position on the sky resulted in a very elongated
synthesized beam; we could not reduce these data completely, the
MIRIAD/restore command failed. This problem did not occur with the combined
ATCA data set. Table~\ref{Tab:obs.summary} lists the  beam size and the
$1~\sigma_{\rm rms}$ error for the combined data set. 

The problem with the October run did not affect a setting of constraints on
the SFR (none of both galaxies was detected).  However, the predicted flux of
a kilonova radio flare  can be rapidly evolving with time; it can be very
different for two different times $dt_1$ and $dt_2$ if $dt_2/dt_1 \sim 3$ 
\citep[see][their Eq. 3 and figure 1]{Fong2016ApJ}. Therefore, in
Fig.~\ref{fig:knlight} we plot only  the constraint on $\nu L_\nu$ for the
first ATCA observing  run  (for a robust parameter of 0.5 the size of the
synthesized beam was $30\farcs4\,\times\,1\farcs8$ and $1\sigma_{\rm
rms}=13.5~\mu$Jy beam$^{-1}$). 

\end{appendix}


\bibliographystyle{apj}
\bibliography{ms.bbl}

\end{document}